# Elastic flexibility, fast-ion conduction, boson and floppy modes in $AgPO_3$-AgI glasses examined in Raman scattering, infrared reflectance, modulated differential scanning calorimetry, ac electrical conductivity and molar volume experiments

**Deassy I. Novita, P. Boolchand**,

Department of Electrical and Computer Engineering, University of Cincinnati, Cincinnati, OH 45221-0030, USA

**M. Malki,**

CNRS, UPR 3079 CEMHTI, 1D avenue de la Recherche Scientifique, 45071 Orléans cedex 2, France and Polytech'Orléans – Université d'Orléans, 8, rue Léonard de Vinci, 45072 Orléans,France

**Matthieu Micoulaut**

Laboratoire de Physique Theorique de la Matiere Condensée, University Pierre et Marie Curie, Boite 121, 4, Place Jussieu, 75252 Paris, Cedex05, France.

## Abstract

Raman scattering, IR reflectance and modulated DSC measurements are performed on specifically prepared dry $(AgI)_x(AgPO_3)_{1-x}$ glasses over a wide range of compositions $0 < x < 60\%$. A reversibility window is observed in the $9.5\% < x < 37.8\%$ range, which fixes the elastically rigid but unstressed regime also known as the Intermediate Phase. Glass compositions at $x < 9.5\%$ are stressed-rigid, while those at $x > 37.8\%$ elastically flexible. Raman optical elasticity power-laws, trends in the nature of the glass transition endotherms corroborate the three elastic phase assignments. Ionic conductivities reveal a step-like increase when glasses become stress-free at $x > x_c(1) = 9.5\%$, and a logarithmic increase in conductivity ( $\sigma \sim (x-x_c(2)^t$ ) once they become flexible at $x > x_c(2) = 37.8$ % with a power-law $t = 1.78$. The power-law is consistent with percolation of 3D filamentary conduction pathways. Traces of water doping lower $T_g$ and narrow the reversibility window, and can also completely collapse it. Ideas on network flexibility promoting ion-conduction are in harmony with the unified approach of Ingram et al., who have emphasized the similarity of process compliance or elasticity relating to ion-transport and structural relaxation in decoupled systems. Boson mode frequency and scattering strength display thresholds that coincide with the two elastic phase boundaries. In particular, the scattering strength of the boson mode increases almost linearly with glass composition x, with a slope that tracks the *floppy mode fraction* as a function of mean coordination number ***r*** predicted by mean-field rigidity theory**.** These data suggest that the excess low frequency vibrations contributing to boson mode in flexible glasses come largely from *floppy modes*.

## I. INTRODUCTION

Solid electrolyte glasses usually consist of alloys of a base oxide- or chalcogenide-glass with a solid electrolyte additive. In many instances the base and additive components form homogeneous alloy glasses over a wide range of compositions [1-4]. In other instances, phase separation occurs as a solid electrolyte phase segregates either on a nanoscale [5] or a microscale [3, 6]. In homogeneous glasses, ions such as $Ag^+$ or $Li^+$ diffuse by hopping and contribute to high ionic conductivity with increasing solid electrolyte content [1-4]. Substantial progress in modeling ion-transport in solid electrolyte glasses has taken place over the past two decades [7-14]. In spite of these developments, we are still not in a position to reliably identify the mechanisms that control fast-ion transport at present so as to optimize the highest conductivity achievable. It is not clear on what aspects of local and intermediate range structures of a glass network, if any, control ion-transport?

It is generally believed [8, 9, 15, 16] that in solid electrolyte glasses physical processes associated with ion transport and network structure are *decoupled*. For example, the ratio ($R_\tau$) of structural ($\tau_S \sim 10^2$ sec) to conductivity ($\tau_\sigma \sim 10^{-10}$ sec) relaxation times near the glass transition temperature ($T_g$) is of the order of $R_\tau \sim 10^{12}$ . On the other hand, in polymer electrolytes such as polyalkylene oxides, $R_\tau$ values of nearly 1 or even less than 1 have been observed [15, 17.] In these *coupled* systems, clearly a reverse circumstance must prevail, viz., ion-hopping and network structure relaxation must be closely tied to each other.

Recently Ingram et al. [18] have investigated the glassy electrolyte system, $(AgI)_{50}(Ag_2O)_{25}(MoO_3)_{25}$ in conductivity and pressure-DSC experiments. The Ag-iodomolybdate is an example of a decoupled solid electrolyte glass system. The activation energy ($\Delta E_A$) and the activated volume ($\Delta V_A$) are estimated from conductivity measurements [19-21]. Pressure-DSC measurements are used to obtain the moduli for structural relaxation $M_s$ from a Pressure dependence measurement of the fictive temperature $T_f$. Ingram et al. suggest that moduli for conductivity relaxation or structural relaxation can be defined as the ratio of activation energy ($\Delta E_A$) and activated volume ($\Delta V_A$)

$$M = \frac{\Delta V_A}{\Delta E_A} \qquad (1)$$

For the case of silver iodomolybdate they find conductivity (σ) and structure (s) relaxation processes, to yield the same moduli, i.e., $M_\sigma = M_s$. They conclude that basic interactions that control ion-transport and network structure relaxation in decoupled (glasses) systems are essentially the same. Ingram et al.[18] identify the ratio M as representing as a "process" modulus, and find decoupled systems like electrolyte glasses possess larger moduli than coupled systems like polymer electrolytes. In glassy electrolytes, the activated volumes, and therefore activated energies are much smaller for ion-transport than for structural relaxation ones, leading to much shorter relaxation times in former processes. These observations are striking and bring *decoupled* and *coupled* electrolytes under a common platform. These ideas suggest that very similar elementary processes are operative over a wide range of relaxation times in electrolytes.

In the present work we have examined the $(AgI)_x(AgPO_3)_{1-x}$ solid electrolyte glass system in a variety of experiments including modulated-DSC, Raman scattering and IR reflectance, AC conductivity and molar volume experiments as a function of glass composition x, paying particular attention to the handling of precursors in synthesis of samples. A brief version of the present work has appeared recently [1]. This particular electrolyte glass has been widely examined in the past by more than a dozen groups [7, 12, 22-33]. Our experiments reveal that traces of water intimately affect network structure, glass transition temperatures [34], network elasticity, ion-transport [1] and, in general, the physical behavior of these glasses much more so than hitherto recognized.[35] In dry samples, we observe the intrinsic elastic behavior of these glasses, which displays three distinct elastic phases [1] as a function of increasing AgI content x; a *stressed-rigid phase* in the 0 < x < 9.5% range, an *intermediate phase* in the 9.5% < x < 37.8% range, and a *flexible phase* in the 37.8 % < x < 54% range. The underlying elastic phase boundaries are quite sensitive to water content of glasses. At higher concentrations of AgI ( x > 54%) glasses segregate into AgI-rich regions. Ionic conductivities of dry samples, examined as a function of increasing AgI concentration, display an increase that occurs in two step-like fashion; one of these steps occurs near x ~ 9.5% once glasses become stress-free, and the second step occurs near x ~ 37.8% when glasses become elastically flexible. In the flexible phase (x > 37.8%), ionic conductivities increase as a power-law as a function of composition. These data suggest existence of characteristic conducting pathways that apparently percolate in 3D once network structure becomes flexible at x > 37.8%. Our findings are in harmony with the notion that rigidity and flexibility of glass network structure is closely related to ion-transport.

The observation of the three elastic phases in a solid electrolyte glass system is a significant finding. Such elastic phases were observed earlier in covalent glasses [36], and in oxide glasses [37]. The observation of these phases in a fast-ion conducting glass, to the best of our knowledge, is a new development in the field. These results suggest that classification of network glasses into three elastic phases based on their connectivity appears to be a generic behavior. It is now feasible to identify the elastic phase a glass composition belongs to by examining the nature of glass transition in modulated-Differential Scanning Calorimetry (m-DSC).

The paper is organized as follows. In section II we present experimental results. In section III we present some theoretical considerations in enumeration of Lagrangian constraints in $AgPO_3$ and AgI. A discussion of the present results follows in section IV. A summary of conclusions appears in Section V.

## II. EXPERIMENTAL

### A. Synthesis

**Sample A**: Dry $(AgPO_3)_{1-x}(AgI)_x$ glasses were synthesized using 99.9%$Ag_3PO_4$ (Alpha Aesar Inc.), 99.5%$P_2O_5$ ( Fischer Scientific Inc.), and 99.99%AgI (Alpha Aesar Inc.) as the starting materials. The bottles containing the starting materials from the supplier were let stand in a nitrogen gas purged glove box [34] ( relative humidity << 1/5%) overnight prior to opening them. The starting materials were weighed in the desired proportion and thoroughly mixed in an Alumina crucible with all handling performed [34] in a glove box (Vacuum Atmospheres model HE-493/MO-5) with a relative humidity much less than 1/5 %. Mixtures were then transferred to a box furnace held at 125°C in chemical hood purged by laboratory air, and heated at 100°C/hr to 700°C. Melts were equilibrated overnight and then quenched over steel plates. Once synthesized glass samples were encapsulated in evacuated ($10^{-7}$ Torr) pyrex tubings for long term storage.

**Sample B**: The principal difference in processing between sample B and sample A was the starting material $Ag_3PO_4$. The bottles containing the starting materials were opened in the laboratory ambient environment for a short time in the early phase of our experiments. This procedure contaminated the starting material by introducing traces of water from laboratory ambient environment. Subsequently the bottles were transferred to the glove bag with a relative humidity ~ 5% and all handling of the materials were performed in there. Other processing steps were identical to the ones used in synthesizing sample A.

**Crystalline $AgPO_3$**: A sample of dry $AgPO_3$ (sample A) was crystallized by heating to 360 $^{O}$C and held at that temperature for 12 hours, followed by cooling to room temperature over approximately 4 hours. The crystalline nature of the sample was confirmed in x-ray diffraction experiments (see below). Molar volume of the crystalline sample was also measured and compared to the starting glass sample (see below).

### B. Molar volumes

Molar volume of samples was established from mass densities established using Archimedes principle. In a measurement, a glass sample piece (100 milligrams or more in size) was mounted on a thin quartz fiber attached to a Mettler Toledo physical balance model B154, and weights recorded in air and in pure ethyl alcohol (AAPER Alcohol and Chemical Co.). A ½ gram piece of Si single crystal, and Ge single crystal was used to calibrate mass density of alcohol. With the set up, we could reliably measure mass density results to an accuracy of 1% or less for glass samples of 100 milligram are more in size.

Molar volumes of dry $(AgPO_3)_{1-X}(AgI)_X$ glass samples A and samples B are summarized in Fig. 1. Also shown in Figure 1 are molar volumes of glass samples reported by Sidebottom [26] and several crystalline standards including dry c-$AgPO_3$, α-AgI and β-AgI [38]. The value of $V_M$ for c-$AgPO_3$ reported in Fig.1 comes from a sample synthesized by crystallizing a $AgPO_3$ glass sample A.

The present data on $V_M(x)$ show that molar volume of the base material, $AgPO_3$, is quite sensitive to water content, with water prone samples (sample B) having a lower molar volume than dry ones. The molar volumes of $(AgPO_3)_{1-X}(AgI)_X$ glass samples steadily decrease with increasing x, but the starting value of $V_m$ at x = 0, controls in a significant way network packing as AgI is steadily alloyed in the base material.

### C. X ray Diffraction

Crystalline $AgPO_3$ sample synthesized in the present work was characterized using a Rigaku Dmax-2100 x-ray diffractometer. The instrument uses a Cu target ($CuK_\alpha$ radiation wavelength of 1.5406 Å). The diffraction data appears in Fig. 2. It reveals the characteristic reflections expected of the polymeric form of crystalline $AgPO_3$.The most intense Bragg peak in XRD results of our sample occurs at $2\Theta = 29.30$ and is found to be shifted to a lower value than the published result ($2\Theta = 31.67$) from the Joint Committee on Powder Diffraction Standards (JCPDS) data file # 11-0640 [39], suggesting that the lattice parameter of our crystalline sample is somewhat larger.

### D. Modulated DSC

A model 2920 m-DSC from TA Instruments was used to study glass transitions. We used a 3°C/min scan rate and 1°C/100sec modulation rate consistently throughout all measurements. Typically about 15 - 20 milligram of a sample in a platelet form was hermetically sealed in Al pans and placed in the head of the calorimeter along with a reference pan. Further details on the method can be found in the earlier reports [6, 34, 37]. The component of the total heat flow (fig.3 (**green**)) that tracks the modulations is called the *reversing* (fig.3 (**blue**)) heat flow. The difference signal between the total and reversing heat flow signals represents the *non-reversing heat flow* signal [40] and it usually displays a Gaussian-like peak (fig.3 (**red**)), as a precursor to the glass transition[40]. The integrated area under the Gaussian profile gives the non-reversing enthalpy ($\Delta H_{nr}$) at $T_g$. The frequency corrected $\Delta H_{nr}$ term was deduced the usual way by subtracting the $\Delta H_{nr}$ term coming down in temperature from the corresponding term going up in temperature[34]. The frequency corrected $\Delta H_{nr}$ term is then independent of the modulated frequency [40]. The $T_g$ value given by MDSC was deduced from the inflexion point of the reversing heat flow step [40]. The quoted value of $T_g$ of a sample is taken as the *mean value* of $T_g$(up) and $T_g$(down) of the reversing heat flow signal, i.e., $T_g$ = ½ ($T_g$(up) + $T_g$(down)). Because of the finite scan rate of 3°C/min used in MDSC scans, there is a small but finite kinetic shift in $T_g$ between $T_g$ (up) and $T_g$ (down) of typically about 2° C in our samples. By averaging, the deduced $T_g$ represents a *scan rate independent* $T_g$. In each instance at least two samples at a given composition were studied from each batch preparation to check for reproducibility of results and insuring homogeneity of samples.

M-DSC scans were performed both on "as quenched or virgin" and "$T_g$-cycled" glass samples [34]. The nature of the m-DSC scans on both "as quenched or virgin" and "$T_g$-cycled" glass samples have been discussed elsewhere [34]. Representative m-DSC scans of present glasses (samples A) at x = 5% (Stressed-Rigid), 17 % (Intermediate) and 40% (Flexible) recorded in the present work are given in Fig. 3. We find that the non-reversing enthalpy at x = 5% (Stressed-rigid) sample displays a broad peak (FWHM, W = 35(2) °C) that is asymmetric (high-T tail), and possesses a frequency corrected integrated area, $\Delta H_{nr}$ = 368.4 (5) mcal/grams. At x = 17%, a composition in the intermediate phase, the width W=25(2) °C, $\Delta H_{nr}$ = 90.6(5) mcal/grams and the peak becomes symmetric. At

x = 50%, a composition in the flexible phase, the width W decreases to 15(2) °C, $\Delta H_{nr}$ = 276.3(5) mcal/grams and the peak remains symmetric. These trends in the width W and shape of the non-reversing heat flow observed in the present glasses are quite similar to those seen earlier in covalent glasses [41], and serve to provide independent confirmation of the assigned elastic phases. We shall return to discuss these results in the next section.

A summary of the present m-DSC results appears in Fig. 4. In panel (a), we summarize variation in $T_g(x)$ for both set of samples A and B. In general we find $T_g$s for both set of samples to steadily decrease as AgI content of glasses increase. For samples of set B, $T_g$s are somewhat lower than in set A largely because these samples contain more water (see section IIA). The present data are compared to two previous reports one by Mangion-Johari [24] and a second one by Sidebottom [26]. In these earlier reports, $T_g$s are also found to steadily decrease as x increases, although the starting value of $T_g$ of the base material ($AgPO_3$) is significantly lower than the value for the present samples. In Fig. 4(b), we plot the observed variation in the non-reversing enthalpy, $\Delta H_{nr}(x)$ for both set of samples, A and B. These data show that the $\Delta H_{nr}(x)$ term for the set of samples A shows a square-well like global minimum in the 9% < x < 38% range [1], the reversibility window. In samples of set B, the reversibility window is found to be narrower, extending to the 20 % < x < 34% range, and to be somewhat shallower in relation to the window in samples A. It is clear that the higher content of water present in samples of set B has a bearing on the reduced width of the reversibility window. Finally, in Fig. 4(c), we plot the variation in heat capacity change at $T_g$, $\Delta C_p(x)$ for samples in set A and B, along with results reported earlier by Hallbruker and Johari [22] on the same glass system. The latter results have come from DSC measurements where one observes only the total heat flow in the scanning calorimetric measurements. It is interesting to note that the variation in $\Delta C_p(x)$ for the three different sets of samples show a steady decrease with increasing AgI content. We will comment on these data later.

### E. Raman scattering

A model T64000 triple monochrometer Raman dispersive system from Horiba Jobin Yvon Inc, equipped with a CCD camera and a microscope (Olympus BX 41 with 80x objective) was used to study Raman scattering of the samples. The scattering was excited with 514.1 nm radiation with the laser beam focus to 2 micron spot size (diameter) using typically 1.2 mWatts of power illuminating a sample. Measurements were performed at room temperature with sample mounted in a MMR Joule-Thompson refrigerator [34] cold stage. Observed lineshapes were analyzed in terms of superposition of suitable number of Gaussians to extract mode-frequency, mode-width and mode-scattering strength intensity. Raman scattering is a powerful light scattering technique used to elucidate molecular structure of glass samples as will be illustrated here.

Figures 5 and 6 give a summary of Raman data on glass samples A in the low ( 0-440 $cm^{-1}$) and high ( 400-1400 $cm^{-1}$) frequency ranges. In the very low frequency range ( < 50 $cm^{-1}$) (Fig. 5), we observe a broad mode usually identified as the boson mode, whose scattering strength intensity steadily increases as a function of AgI content of glasses. At x > 37.8%, this particular mode becomes the most intense feature of the observed line shape in glasses. These features of our data are generally similar to those reported earlier by Fontana et al.[42]. In our experiments, we have normalized Raman spectra at different compositions to the <u>same</u> laser power, and as a check of this normalization find mode scattering strength intensities in the bond-stretching regime to change systematically with x. Raman scattering is in general given as [43-45],

$$I_{\exp} = \frac{C(\omega)g(\omega)[n_B + 1]}{\omega} \qquad (1)$$

In equation (1), C(ω) represents the photon-vibration coupling constant [46], g(ω) the vibrational density of states (VDOS), and $n_B$ the Bose occupation number. To analyze the low frequency (< 200 $cm^{-1}$) modes in glasses, we obtained the reduced Raman scattering ( $I_{red}$) from $I_{exp}$,

$$I_{red} = \frac{I_{\exp}}{n_B + 1} \propto \frac{C(\omega)g(\omega)}{\omega} \qquad (2)$$

by dividing $I_{expt}$ by the factor ($n_B$ + 1). Since C(ω) varies [47] as ~ ω at frequencies ω > 20 $cm^{-1}$ , the ratio $\frac{I_{\exp}}{n_B + 1}$ provides a reasonable measure of the vibrational density of states g(ω). Reduced Raman scattering at select glass compositions appear in Figure 7. The effect of correcting for the finite T by obtaining $I_{red}$ now yields lineshapes that show the low frequency modes to display symmetric peak-like features. The low frequency modes were analyzed in terms of three modes, one centered near 40 $cm^{-1}$ (related to boson mode), a $Ag^+$ restrahlen mode near 85 $cm^{-1}$, and a mode of AgI near 130 $cm^{-1}$ (Figure 7) [48-50] .Variations in the frequency and scattering strength reduced Raman intensity of the boson mode reveal thresholds near the stress ($x_c(1)$ = 9.5%) and rigidity ($x = x_c(2)$ = 37.8%) transition (Figure 8), the two elastic thresholds observed in these glasses. The boson mode scattering strength intensity (Figure 8b) shows a small step-like increase near $x_c(1)$ , and a rather striking linear increase at $x > x_c(2)$. A linear extrapolation of the $I_{red}(x)$ data from x = 37.8% to x = 100% yields, $I_{red}(x = 100) = 380$ counts, and permits to define a normalized scattering strength $\frac{I_{red}(x)}{I_{red}(100)}$, and a slope $\frac{d\left[\frac{I_{red}(x)}{I_{red}(100)}\right]}{dx}$. The linear behavior is characterized by a slope of 1.49, and we shall return to discuss these data in section IV E.

At a slightly higher frequency (figures 5 and 6), we then observe *bond-stretching modes* associated with different structural units [48-50] ( 300- 650 $cm^{-1}$ range). In the high frequency region (Fig. 6), we then observe *bond-stretching modes* of $PO_4$ units present in *chains* that steadily red-shift and decrease in strength, as modes of $PO_4$ units in *small rings* and *large rings* steadily increase in scattering strength with increasing x. We shall return to discuss the mode assignments after presenting the IR reflectance data.

How do the present Raman data compare to previous results in the field?

Raman scattering in $(AgPO_3)_{1-X}(AgI)_X$ glass samples of set B (less dry) excited with 514.1 nm radiation show strong fluorescence compared to set A, and the lineshapes did not change much with glass composition. Similar results were obtained by Shastry and Rao [51] at glass compositions of x = 0.3, 0.4 and 0.5, with $T_g$s respectively of 109°C, 95°C and 87°C. These $T_g$s are similar to those reported by Mangion and Johari [23] (Fig. 4a). Using 488 nm excitation, Shastry and Rao [51] were able to suppress fluorescence and obtained good Raman signals but their data also show little change in lineshape with AgI content. These differences of Raman lineshapes between present results and earlier reports suggest that the molecular structure and thus the vibrational density of states of driest (high-$T_g$) and less dry (low-$T_g$) samples are quite different. We will return to discuss these data in section IV.

## F. Infrared reflectance

Fourier Transform-Infrared (FTIR) spectra were collected at room temperature using a Thermo Nicolet Nexus 870 FTIR bench. Specular reflectance measurements were made using a Seagull accessory, and a high reflectivity low carbon stainless steel mirror was used as reference to normalize the reflectivity signal. Each spectrum was collected for 200 scans at 8 $cm^{-1}$ resolution. A proper combination of source, beam splitters, and detectors allowed collecting spectra in the 50-4000 $cm^{-1}$ range. To obtain signals characteristic of our glass samples, we chose reasonable thickness of bulk samples with flat areas and gave no surface treatment prior to recording infrared reflectance. In these specular reflectance measurements the reflectance signal comes from sample surface. Surface treatment such as polishing was therefore avoided so as not to change the surface structure of glasses[52]. Reflectivity data was then analyzed by Kramers-Kroning analysis to extract absorption signals. Figure 9 shows absorption signals at select glass compositions (the same ones for which Raman data are presented in Figure 6). These absorption signals were then least squares fit to a superposition of Gaussian profiles to extract mode-frequency, intensity and width.

How do the infrared absorption data change as a function of AgI content? A number of observations can be made. First, a perusal of Figure 9 shows that the asymmetric P-$O_t$ vibration mode in long chains systematically red-shifts with AgI concentration. These IR results complement the Raman results of a red shift of the symmetric P-$O_t$ mode of long chains seen earlier in Figure 6. Second, the intensity of the asymmetric mode of terminal $PO_3^{2-}$ groups is found to increase as the AgI content of the glasses increases (figure 9). The behavior is reminiscent of the corresponding symmetric vibrations in the Raman results (figure 6).Third, IR active modes characteristic of rings (970 & 900 $cm^{-1}$) grow at the expense of those of long chains ( 1300 cm-1) as AgI content increases. Raman active modes of rings (near 1000 & 700 cm-1) grow at the expense of modes of long chains (near 1100 & 680 $cm^{-1}$). These observations taken together reveal that glass structure is steadily getting less connected as AgI is introduced in the base $AgPO_3$. Wakamura et al.[53], and Bhattacharya et al.[54] have also reported on IR results on this glass system, and our results differ from these earlier reports. These differences come from the sample make up and also in the way these samples were measured. We believe our samples are dryer than these earlier reports. Furthermore, in their experiments, Bhattacharya et al.[54] performed transmission measured using platelets of intimately mixed finely powdered glass samples with CsI, an IR transmitting matrix.

### G. Raman and infrared vibrational mode assignments

The complementary role of Raman scattering and Infrared reflectance serves as a powerful diagnostic tool for mode assignments. In general, for a centrosymmetric molecule such as $PO_4$ tetrahedral units, no Raman-active vibration is also infrared active and no infrared active vibration is also Raman active. The exclusion rule [55] is quite useful in mode assignments. We shall see that symmetric vibrational modes are strongly excited in Raman scattering but weakly in IR reflectance, on the other hand asymmetric modes are strongly excited in IR reflectance but weakly in Raman scattering. The power of IR reflectance as a probe of both *free* water [56, 57] and bonded water as the OH stretch vibrations of P-$OH^-$ bonds [56, 58] has been widely recognized. In a previous contribution we illustrated the idea for the case of $AgPO_3$. [34]

**Mode assignments**: In comparing the Raman and IR reflectance data on AgI-$AgPO_3$ glasses (Fig.6 and 9) the complementary nature of activity of certain vibrational modes becomes apparent. These features of the present data along with previous work in the field, has permitted making mode assignments. **P-$O_t$ modes**: The highest frequency modes (1235 $cm^{-1}$and 1140 $cm^{-1}$) observed can be identified respectively with the asymmetric and symmetric modes of P-$O_t$ bonds of $PO_4$ terahedra. The asymmetric mode (1235 $cm^{-1}$) is strongly excited in IR but weakly in Raman scattering. The reverse is the case, for the symmetric vibration, as expected. **Defect $(PO_3)^{2-}$** : The next two lower frequency modes, one near 1087 $cm^{-1}$ and the other near 1029 $cm^{-1}$, represent respectively the asymmetric and symmetric vibrations of a $(PO_3)^{2-}$ unit that form part of pyrophosphate grouping of $Ag_2P_2O_7$ unit.[48, 49, 59] In this dimeric unit, two $PO_4$ tetrahedra share a bridging oxygen, and the mode in question represents the P-$O_t$ vibrations of these pair of $Q^1$ species (in NMR notation). The asymmetric (1087 $cm^{-1}$) mode of $PO_3^{2-}$units is strongly excited in IR but weakly in Raman scattering. On the other hand, the reverse is the case for the symmetric mode (1029 $cm^{-1}$) of these units (Table 1). Our data shows that the concentration of these dimeric units steadily increases as AgI content of glasses increases (Figure 9). **Long-chain P-$O_{br}$ modes**: we identify the modes near 686 $cm^{-1}$ and 900 $cm^{-1}$ respectively with the symmetric and asymmetric vibrations P-$O_{br}$ bonds present in long chains (Table 1) of $PO_4$ units. The symmetric vibration is strongly excited in Raman scattering (Figure 6) but weakly in IR. The reverse circumstance prevails for the asymmetric stretch of the P-$O_{br}$ bonds in long chains. The higher vibrational mode frequencies of P-$O_t$ modes in relation to P-$O_{br}$ ones are consistent with the smaller P-$O_t$ bond length (145 pm) than P-$O_b$ bond lengths (160 pm) [60.] **Large ring modes:** As long chains transform into large rings and smaller rings, one expects the vibrational mode frequency of the P-$O_{br}$ bonds to up shift in frequency as a lowering of the P-$O_{br}$-P bridging oxygen bond angle [59] must occur to require closure of small rings. We identify the modes near 750 $cm^{-1}$ and near 920 $cm^{-1}$ with the symmetric and asymmetric vibrations of large rings. The feature near 750 $cm^{-1}$ is actually split into two modes, 733 $cm^{-1}$ and 766 $cm^{-1}$, and we note that both these modes are strongly excited in Raman while their asymmetric counterpart are strongly excited in IR (figure 9). **Small ring modes**: The complementary activity of the two vibrational modes near 970 $cm^{-1}$ and 1000 $cm^{-1}$ in Raman and IR response suggests that these modes are the asymmetric and symmetric vibrations of small rings. In Table 1 we have summarized the present assignments along with suitable references to previous work in the field that broadly support of our findings.

### H. Raman optical elastic power-laws

The narrow vibrational mode associated with symmetric stretch of P-$O_t$ bonds in Raman scattering of samples A (Figure 6) is reasonably well resolved. This particular mode is of interest because it describes the vibrational motion of all $PO_4$ tetrahedra in the structure of the glasses. These tetrahedra are cross-linked across $Ag^+$ cations and are vibrationally coupled. With increasing AgI content, the mode softens and tracks the elastic softening of the backbone. A parallel behavior of a softening of the P-$O_t$ mode in Raman scattering of the present glass system was also observed by Fontana et al. [42]. With fewer compositions studied over the glass forming range, these authors however could not recognize evidence of two thresholds, one near 15% and the other near 40% of AgI (Fig.2 of ref of Fontana et al.[42]) present in their data as well. We have analyzed the red-shift of the mode quantitatively by least-squares fitting the observed lineshapes to extract the width, frequency and integrated intensity of the modes in question. Figures 10a and 10b give a summary of the results on the variation of the P-$O_t$ symmetric mode frequency in Raman experiments, and separately the P-$O_t$ asymmetric mode frequency in IR absorption experiments as a function of AgI. We observe the P-$O_t$ symmetric mode to systematically red-shift as AgI content of the glasses increases.

In both Raman and IR response, we find that the observed variation in mode frequencies displays two vibrational thresholds [1], one near x = 9.5%, and the other near x = 37.8%. These thresholds coincide with the walls of the reversibility window (Fig.4(b)). The frequency variation of the symmetric P-$O_t$ mode observed in Raman scattering can be analyzed further to extract underlying optical elastic power-laws in the Intermediate phase and in the stressed rigid phase as we illustrate next.

In Figure 10a, the curvature in the $\upsilon(x)$ plot of the symmetric stretch mode frequency of P-$O_t$ bonds in the stressed-rigid and in the Intermediate phase can be analyzed by fitting the observed mode frequency squared $\upsilon^2(x)$ to the following equation,

$$\upsilon(x)^2 - \upsilon_c^{\,2}(1) = A[x - x_c(1)]^{p_1} \tag{3}$$

where $\upsilon_c$ (1) represents the threshold value of $\upsilon(x)$ at $x = x_c(1) = 9.5\%$ , the stress transition [1], and $p_1$ the elastic power-law. The optical elastic power-law in the stressed-rigid phase, $p_1$ = 1.25 (2), is thus determined from the slope of a log-log plot shown in

Figure 11a. A similar procedure was used to extract the elastic power-law in the intermediate phase as illustrated in Figure 11b, and a power-law, $p_2 = 0.98$ (3) is obtained using the rigidity phase boundary $x_c(2) = 37.8\%$. We shall return to discuss these results in the next section.

## I. AC electrical conductivity

Glass sample disks about 10mm in diameter and 2 mm thick were syntesized by pouring melts in special troughs. Platelets were then thermally relaxed by cycling through $T_g$. Platelets then were polished, and the Pt electrodes deposited. Platelets were spring loaded in a sample holder, and AC conductivity measurements were taken using an Impedance spectrometer (Solartron SI 1260) as a function of temperature in the 200K < T < $T_g$ range, and as a function of frequency in the 1 Hz < f < $10^6$ Hz range. The room temperature conductivities for each glass samples were deduced from AC electrical conductivity measurements at low frequencies when f approaching 0. The Activation energies ($E_A$) for electrical conductivity for all of the glass samples were extracted from the Arrhenius plot of the dc electrical conductivity as a function of temperature. Further details on the AC electrical conductivity measurement used in present work is described elsewhere [1.] The Coulomb energy for ion creation was estimated in the usual way [1], and the resulting strain energy $E_s = E_a - E_c$ was then obtained. Figure 10 shows a plot of the compositional trends in conductivity $\sigma(x)$ along with $E_a(x)$, $E_c(x)$, $E_s(x)$ data on the present electrolyte glasses.

How do the present results on $\sigma(x)$ compare with earlier ones on the present electrolyte glasses? In Fig. 13b we show compositional trends in $\sigma(x)$ from several earlier studies including those of Bhattacharya et al. [54], Sidebottom [26], Mangion and Johari [24], and Malugani et al. [61], along with present findings in Fig.13a. The σ(x) results of Malugani et al. [61] and Sidebottom [26] agree with each other over a wide range of compositions. On the other hand, the σ(x) data of Bhattacharya et al. [54] are nearly two orders of magnitude higher than those of Malugani et al. [61] The σ(x) data of Mangion and Johari [24] are quite similar to those of Malugani et al. [61] at x > 40%, but at lower x, their data lies in between those of Bhattacharya et al. [54] and Sidebottom [26]. These earlier data are in sharp contrast to the variation of σ(x) reported by us which displays two distinct thresholds. The conductivity of the base $AgPO_3$ glass (x = 0), reported by various groups differ substantially from each other (Figure 11b), and we have suggested that variability of these data is due to presence of water impurities.[34] Once water is incorporated in the base material, it transforms from being elastically stressed-rigid to flexible, and one no longer expects the conductivity to display the two thresholds as will be discussed later.

## III. LANGRANGIAN CONSTRAINTS IN $AgPO_3$ GLASS AND AgI GLASS

AgI exists in an orthorhombic phase (β) at room temperature. At T > 150°C it transforms to a cubic (α) solid electrolyte phase. In this phase Ag migrates across tetrahedral and trigonal interstial sites resulting in high ionic conductivity ($10^{-1}$ $ohm^{-1}cm^{-1}$). In an earlier contribution [3], the Lagrangian constraints associated with the α-phase were estimated assuming that only bond-stretching forces were intact for both Ag and I. Since Ag is mobile, the bond-bending forces centered on Ag and on I must be weak, and thus assumed to be intrinsically broken. Such a model of the α-phase is a good starting point to describe the glass, which would be the case only if the glass were a fast-ion conductor.[3] These estimates place the constraints/per atom, $n_c$ = 2, for AgI glass, suggesting a mean coordination number $\boldsymbol{r}$ = 2. AgI glass is thus expected to be an elastically flexible system. The glass transition temperature of AgI is also expected to be low (~ 50°C) since $\boldsymbol{r}$ = 2, as is indeed observed when AgI is alloyed in base chalcogenide glasses.[3] In this context, a model glass to compare AgI glass would be Se ( $\boldsymbol{r}$ = 2) with a $T_g$ of 40°C.

Crystalline $AgPO_3$ is composed of $PO_4$ terahedra arranged in helical chains.[62] In a helical pitch there are 4 tetrahedral units. In each $PO_4$ tetrahedral unit, a Phosphorus atom possesses a coordination number ($\boldsymbol{r}$) of 4 with two bridging ($O_b$) and two terminal ($O_t$) oxygen atoms. Ag sits between these chains in a distorted tetragonal pyramid having a nominal coordination number of 5. However, two of these 5 Ag-O bonds are short ( ~ 2.40 A) and the other three bonds are somewhat longer (~ 2.50 A).[62] The two short Ag-O bonds are to terminal oxygen atoms in a chain, as one would have expected with $Ag^+$ compensating for the two terminal $O^-$ atoms. In the glassy phase, the simplest assumption is that the near-neighbor bonding constraints with the 2 $O_t$ are strong and intact while those with the 3 more distant bridging Oxygen atoms are weaker and intrinsically broken. In this *limiting case*, the coordination numbers of P, $O_b$, $O_t$ and Ag become respectively 4,2,2,2. An atom with coordination $\boldsymbol{r}$ (> 2) , has $\boldsymbol{r}/2$ bond-stretching and $2\boldsymbol{r}$-3 bond-bending forces. The Lagrangian constraints for P, $O_b$, $O_t$, Ag are then respectively 7,2,2,2. For a $AgPO_3$ formula unit (N = 5), we enumerate the Lagrangian constraints as,

$$5n_c = \left[ 7(P) + \frac{1}{2} \times 2 \times 2(O_b) + 4(O_t) + 2(A_g) \right] = 15 \qquad (4a)$$

$$\text{or} \quad n_c = 3 \qquad (4b)$$

In the *limiting case*, $AgPO_3$ glass is thus viewed to be an example of an isostatically rigid network. In practise, this appears nearly to be the case. Our experiments show $AgPO_3$ glass to be weakly *stressed rigid* glass ( $n_c > 3$), we suppose because the Ag residual constraints with the 3 more distant oxygen near-neighbors cannot be totally excluded. However, upon alloying a mere 9 molar % of AgI, interchain-spacing increase enough to lift the residual constraints, and an isostatically rigid alloyed glass is realized at the onset of the reversibility window (fig.2b). The coordination number decrease of Ag with O is independently corroborated by coherent X-ray scattering results [63] on these glasses that show a decrease from 5.1 at x = 0 to 2.5 at x = 0.5.

In summary, these considerations on Lagrangian constraints suggest that $AgPO_3$ glass can be expected to be mildly stressed-rigid and AgI glass to be elastically flexible. In the present solid electrolyte glasses, $AgPO_3$-AgI, the underlying networks should steadily soften as the AgI content of these glasses is steadily increased.

## IV.DISCUSSION

### A. Molecular structure of glasses

$AgPO_3$ glass is widely regarded [35, 60, 62, 64] to be made up of polymeric chains of $PO_4$ tetrahedral units, and glass structure has often been compared to one of its crystalline [35, 62] polymorphs, which is made up of helical chains of $PO_4$ units. Raman scattering from $AgPO_3$ glass and its polymeric crystalline counterpart is dominated by two vibrational modes, one due to the symmetric stretch of P-$O_t$ bonds ( 1140 $cm^{-1}$) and the other due to symmetric stretch of P-$O_{br}$ bonds ( 686 $cm^{-1}$)[34, 48-51, 59, 60, 65-67]. In the crystalline phase these modes are rather narrow (linewidths ~ 6 $cm^{-1}$) but in the glass these modes are understandably broad (linewidth ~ 40 $cm^{-1}$) [34] due to intrinsic disorder. One also observes weaker features in Raman vibrational density of states of $AgPO_3$ glass that can be identified to the characteristic symmetric and asymmetric P-$O_{br}$ vibration of large rings near 766-733 $cm^{-1}$ and 920$cm^{-1}$ observed respectively in Raman and IR response. The corresponding vibrational features of small rings include the symmetric and asymmetric stretch of P-$O_{br}$ bonds near 1000 $cm^{-1}$ and 970 $cm^{-1}$ observed respectively in Raman and IR response (Table 1, fig. 6 & 9). This is not all that surprising given that crystalline $AgPO_3$ also exists in a tetrameric form composed of 4 membered rings of $PO_4$ tetrahedra. In addition, one observes Raman vibrational density in the 300-650 $cm^{-1}$ range that can be identified to the bending modes of different structural $PO_4$ units (Table 1, fig. 5-6).

Upon comparing Raman with the IR in glasses one finds that vibrational features attributed to large and small rings steadily increase in integrated intensity at the expense of those of polymeric chains. For example, in Raman scattering the symmetric stretch vibrations of P-$O_{br}$ bonds in small rings ~ 1000 $cm^{-1}$ and large rings ( ~766-733 $cm^{-1}$) far exceed the intergrated intensity of the corresponding vibration in long chains ( 680 $cm^{-1}$) as x ~ 50%. A similar pattern is noted in the IR absorption data shown in Figure 9. One finds for example the asymmetric P-$O_{br}$ mode in large rings ~ 920 $cm^{-1}$ and small rings near 970 $cm^{-1}$ to systematically evolve in strength as the corresponding P-$O_{br}$ mode in long chains near 900 $cm^{-1}$ steadily decreases as AgI content of glasses increases to 50%. These results lead us to believe that with increasing concentration of AgI, glass structure steadily evolves from polymeric chain-like at low x to becoming ring-like as x increases to 50%.

The structure of the present glasses has been examined in neutron diffraction [27] and x-ray diffraction experiments, which reveal a First Sharp Diffraction Peak (FSDP) at anomalously low momentum transfer Q ~ 0.7-0.8 $cm^{-1}$. The FSDP is widely regarded as signature of Intermediate Range order in glasses. Reverse Monte Carlo modelling of the neutron and x-ray structure factors by Wicks et al. [27] and by Borjesson et al.[12] have suggested that the FSDP is related to the increased inter-chain spacing of polymeric phosphate chains as AgI is alloyed. The insertion of the AgI additive between polymeric chains drives them apart and shifts the FSDP to lower Q values in the experiments as confirmed by the calculations. Alternative suggestions [28, 64, 68] of the FSDP in terms of segregation of AgI-rich clusters in the glasses are not only inconsistent with the x-ray and neutron structure factors, but also with the existence of a solitary glass transition that steadily decreases with AgI content (Figure 4).

The broad picture of structure described above appears to be supported by the present Raman results at low AgI additive, x > 25%. In figures 6 , the vibrational mode near 1140 $cm^{-1}$, ascribed to $PO_t$ - $Ag^+$ chain segments, is found to steadily soften ( Figure 10a). We can understand a weakening of the intra-chain P-$O_t$ bonding interactions at the expense of inter-chain $PO_t$-$Ag^+$ ones as more $Ag^+$ cations become available between the chains. However, as x > 25%, a better description of the glasses consists of visualizing them to be made of large- and small-rings of $PO_4$ tetrahedra and fragmented $Ag_2P_2O_7$ pyrophosphate units rather than polymeric chains. This is not unexpected given that the base glass, $AgPO_3$, is also known to exist in a crystalline polymorph that consists of tetramers. Rings pack better than chains and probably account for the decreasing molar volume (Figure 1) as the AgI content of the glasses increase. The loss of global connectivity associated with the chain to ring transition of the network structure leads to two elastic phase transitions from mildly rigid to flexible glasses as x increases to 54%.

### B. Three elastic phases in $(AgPO_3)_{1-x}(AgI)_x$ glasses

The central finding of the present work is the existence of three distinct elastic regimes in the present electrolyte glasses as their connectivity is lowered by alloying the electrolyte salt. Several pieces of experimental data bear on the subject including (i) the non-reversing enthalpy at $T_g$, (ii) Raman optical elastic power-laws and (iii) nature of the glass transition endotherms.

**Non-reversing enthalpy at $T_g$**: The non-reversing enthalpy at $T_g$, $\Delta H_{nr}(x)$, displays a wide, deep and sharp reversibility window in the present glasses (Figure 4b). Such windows were observed earlier in covalent glasses [69], and were identified with networks that belong to a phase that is elastically rigid but unstressed, also known as the Intermediate Phase (IP). In analogy to those findings, we thus identify the compositional window, 9.5 % < x < 37.8%, (Figure 4b) with the IP in the present electrolytes. In covalent glasses atoms usually bond in conformity of the 8-N coordination rule facilitating estimates of network connectivity by a count of the mean coordination number ***r***. It is usually defined as

$$r = \left(\frac{1}{N}\right)\Sigma n_i r_i \qquad (5)$$

Where $\left(\frac{n_i}{N}\right)$ and $r_i$ designate respectively the fraction of atoms 'i' possessing a cordination number $r_i$. In the present electrolytes atom valences, and particularly that of the mobile Ag , is not satisfied locally [13] and estimates of network connectivity pose new challenges. However, we note that the connectivity of the end member glass compositions, viz., $AgPO_3$ ( $\boldsymbol{r} > 2.4$) and AgI ( $\boldsymbol{r} = 2$) have been estimated using constraint counting algorithms (section III), and these data unequivocally suggest that the mean coordination number, $\boldsymbol{r}$, (or network connectivity) of the present glasses must steadily decrease as x increases. That view is independently corroborated by the measured glass transition temperatures $T_g(x)$, that steadily decrease as x increases (Figure 4a). Stochastic agglomeration theory [70] has shown that $T_g$ provides a reliable measure of network connectivity. These considerations suggest that glass compositions at low x, $0 < x < 9.5\%$, must belong to the *stressed-rigid phase*, those at high $x > 37.8\%$ to the flexible phase, while compositions at intermediate values of x ($9.5\% < x < 37.8\ \%$) to the Intermediate phase. These assignments are corroborated by Raman optical elastic power-laws as we discuss next.

**Raman optical elastic power-laws**: Raman scattering experiments in covalent systems has proved to be particularly diagnostic in identifying the different elastic regimes of network glasses through measurements of optical elastic power-laws. The conditions that have made Raman scattering as a rewarding probe of elastic phases in glasses was discussed in a recent review [36] .Numerical experiments on stochastic amorphous Si networks modelled by a Keating potential, i.e., bond-stretching and bond-bending forces, reveal elastic constants ($C_{11}$) to increase as a power-law [71, 72] in the stressed-rigid regime,

$$C_{11} \approx (r - r_c)^p \qquad (6)$$

with p ~ 1.4. In these simulations networks of different connectivity ***r*** are made by cutting bonds. Here $\boldsymbol{r}_c$ represents the phase boundary between flexible and stressed-rigid elastic phases.

In the Raman scattering experiments a select vibrational mode is chosen and the square of the mode frequency (proportional to optical elasticity) is studied as a function of glass composition (network connectivity). The select vibrational mode in question must belong to a local structure that forms part of the connective tissue and is usually close to being optimally constrained. In the group IV-chalcogenides, corner-sharing (Ge or Si)$(Se_{1/2})_4$ tetrahedra that are strongly Raman active, have been particularly useful probes[73, 74, 36] of elastic behavior in the chalcogenide glasses.

In the present electrolyte glasses the symmetric stretch of $P\text{-}O_t$ bonds in $PO_4$ tetrahedra give rise to a strongly excited Raman band near 1140 $cm^{-1}$. This band encompasses contributions from $PO_4$ tetrahedra present in long-chains, large-rings and small-rings. In section II H, we had deduced from analysis of the Raman data the elastic power-law in the stressed rigid phase, $p_1$ = 1.25 (3), and in the intermediate phase, $p_2$ = 0.98 (3). These power-laws, surprisingly, are found to be remarkably similar to the ones reported earlier in covalent systems. For example, in ternary $Ge_xP_xSe_{1-2x}$ glasses [36], the elastic power-law in the stressed-rigid phase, $p_2$ =1.48(2) , and in the intermediate Phase, $p_1$ = 0.98(1). Parallel results are found in other covalent systems[3, 73]. To summarize, the measured Raman elastic power-laws in the present electrolyte glass strongly support the suggested identification of the three elastic phases from the non-reversing enthalpy at $T_g$.

**Nature of glass transition endotherms**: In the chalcogenide glasses we have observed that the non-reversing heat flow in the three elastic phases displays characteristic trends. In the flexible phase, the heat flow terms usually displays a narrow (~ 20 $^{o}$C) and symmetric temperature profile, with the enthalpy slowly increasing as glasses age. In the Intermediate phase the heat flow term displays a minuscule enthalpy that shows little or no aging. In the stressed-rigid phase the non-reversing heat flow term is usually broad and displays an asymmetric temperature profile with the underlying enthalpy steadily increasing upon aging of glasses. These characteristics of the non-reversing heat flow in covalent systems have close parallels to those in the present electrolyte glasses (section II D). Although we have not been able to study aging effects in the present electrolytes, we observe the non-reversing heat flow endotherm to be wide and asymmetric at $x < 9.5\%$ , and to be narrow and symmetric at $x > 37.8\%$, trends that follow the ones noted in chalcogenides earlier.

### C. Role of water impurities on physical behavior of $AgPO_3$-AgI glasses

The samples investigated in the present work differ from ones reported by earlier groups [24, 26, 54] by virtue of synthesis, and we believe our samples are dryer and contain less water. How does that influence of physical properties and structure of glasses? It may be useful to begin the discussion by reviewing results on present samples, particularly comparing results on set A (driest) with set B (less dry) . If one compares $T_g$ and $V_m$ data one finds a distinct pattern; $T_g$s of samples in set B are lower than those of set A by anywhere from about 40°C to 10°C (Figure 4a), and molar volumes of samples in set B are lower than those of set A by anywhere from 4.5% to 0.5% (figure 1). Likewise, $T_g$s of samples reported by Sidebottom are 60°C lower than those found in samples of set A (Figure 4a) while molar volumes ($V_m$) of his samples are anywhere from 4.76% ( x = 0) to 11. 8% ( x = 50%) lower that the ones observed by us for samples of set A. $T_g$s of samples reported by Mangion and Johari are anywhere from 100°C to 60°C lower than those of samples in set A. These data suggest that presence of water in samples lowers $T_g$ and molar volumes [58, 75], and handling of precursors in a dry ambient is necessary to minimize water uptake by samples [34, 35].

The underlying structure issues bears on chains of - P-O-P- steadily depolymerizing by presence of water impurities for which evidence comes from IR absorption experiments that reveal features [34, 56] characteristic of P-$OH^-$ vibrations near 1637 $cm^{-1}$, 2318 $cm^{-1}$, and 2352 $cm^{-1}$. With increasing AgI content of present glasses these vibrational features reduce in strength but never vanish even for the set of samples A. Presence of depolymerised chains lower the connectivity of the network that is reflected in the $T_g$s of the samples. Depolymerized chains apparently pack better that is clearly reflected in molar volumes of glasses.

A significant finding of the present work is the strong (factor of three or more) enhancement of the mid-IR response of glasses with increasing AgI content, a feature that is readily seen in comparing response at x = 0 with the one at x = 50% , both in the bond-stretching and bond-bending regimes (figure 9). The observation is reminiscent of the mid-IR response enhancement (at least by factor of 4) in wet $AgPO_3$ glasses in relation to dry ones [34] which was traced to collective modes of water locking on to the $Ag^+$ floppy modes and lead to a long range coherent enhancement of oscillator strength of all optic modes of chains. A similar circumstance, most likely, exists in the present dry glasses with AgI salt (playing the same role as that of water in) dressing the surfaces of the smaller structural motifs such as small rings and pyrophosphate groupings and transferring the collective modes of AgI salt by locking on to $Ag^+$ floppy modes and lead to long range coherent enhancement of optic modes.

The very striking role played by traces of water impurities in the narrowing of the reversibility window [37] in going from the set A to the set B of samples is illustrated in Figure 4b. We find, for example, that $x_c(1)$ shifts up from 9.5 % to 22%, while $x_c(2)$ shifts down from 37.8% to 35%, thus narrowing the Intermediate phase width from $\Delta x = x_c(2)-x_c(1) =$ 28% in set A to nearly 13% in set B of samples. IR response of these glasses reveals largely bonded water with little or no evidence of free water in samples. Replacement of bridging oxygen atoms in chains by dangling $OH^-$ ends serves to splice the P-O-P chain network, and reduce the range of glass compositions across which a stress-free network can persist. These findings are reminiscent of the collapse of Intermediate phases of chalcogenide glasses ($GeSe_4$ and $GeS_4$) when alloyed by iodine [76]. Halogen atoms in the chalcogenides serve to replace bridging chalcogen (S,or Se) atoms and disrupt the network structure by creating dangling Ge-I ends [76]. At present we are uncertain on aspects of local or intermediate range structures that control the width of the IP in the present electrolyte glasses, and this is a point of continuing investigations.

### D. Ionic conductivity and Network Flexibility

The AC conductivity results on the present electrolyte glass system raise two obvious questions, how does one reconcile the two step-like variation in $\sigma(x)$ observed here with earlier work in the field? Secondly, what is the microscopic origin of these two steps in σ(x)? We begin by addressing these issues. $T_g$ endotherms of dry $AgPO_3$ glasses are characteristic of stressed-rigid networks [34], while those of wet samples characteristic of flexible networks. Wet samples possess not only lower $T_g$s but also have a non-reversing heat flow term [34] that is narrow and symmetric in T, features that we have identified [41, 77] earlier with flexible glasses. Once the base glass has been rendered flexible by water doping one does not expect to observe the stress and rigidity transitions and their consequences on conductivity (See section II. I & IV. C). Thus, the two thresholds observed in σ(x) in the present work are characteristic of dry samples and represent the intrinsic behavior of the electrolyte glasses. These were not observed in earlier reports largely because samples reported upon in the literature, to the best of our knowledge, are wet samples in which the base oxide ($AgPO_3$) is already flexible.

The small increase in conductivity near x ~ 9.5% can be traced to the small decrease in strain activation energy $\Delta E_s$ (Fig. 12) as glasses become stress-free at x > 9.5%. In this range of compositions, we visualize glass structure to be largely composed of polymeric $PO_4$ chains (Figures 6 and 9) that become separated as AgI is inserted between them. Such a simple structural description was in fact, proposed many years ago by Wicks et al.[27] as mentioned earlier. These hard sphere Monte Carlo simulations also showed evidence of $Ag^+$ diffusivities increasing precipitously at x > 25%. This composition is not that far from the composition, $x_c(1)$ = 9.5% observed in our experiments. We attribute the increase in conductivity of glasses at x > 9.5% to result from the weakly alloyed chain glass structure becoming stress-free.

The real structure of these electrolyte glasses especially at higher AgI concentrations is likely to be different than the one described by the Reverse Monte Carlo simulations[12, 27]. The present Raman and IR data reveal that there exists a transition from a chainlike to a ringlike structure as the AgI content x > 40% and glasses soften and become elastically flexible at $x > x_c(2) = 37.8\%$. The exponential increase of conductivity in the flexible phase give rises to a power law [1] , $\mu = 1.78$ (10) (figure 12). This value is in excellent agreement with the predicted power-law ($\mu = 2.0$) of conductivity percolation in 3D resistor networks [78]. These data suggest that there must exist in the electrolyte glasses percolation paths for $Ag^+$ ions to hop along. At low x these paths must be isolated or decoupled. But as the concentration of these paths increases with x, these paths must percolate and contribute to the logarithmetic increase of conductivity once x > 37.8%. To gain better insights into these conductivity thresholds more sophisticated structural models of these glasses have to be constructed that can at least predict the Intermediate Phase reported here. It would then, realistically, be possible to model the second conductivity transition near $x = x_c(2)$.

Experimental evidence for filamentary nature of ionic conduction in the present electrolyte glasses at x = 50% (flexible phase) have come from field ion-microscopy measurements of Escher et al.[79.] These authors examined emission from tips of wires made of $(AgI)_{50}(AgPO_3)_{50}$ . In their experiments they imaged tips as a function of time and observed several bright spots identified with $Ag^+$ ion emission centers. These centers are thought to arise from nanometer-sized ion channels terminating on tips. These field-ion measurements confirm in a direct way the conductivity enhancement predicted by the 3D percolation behaviour in the present electrolyte glasses.

The broader issue raised by the present results bears on the fundamental interactions that determine fast-ion transport and network structural interactions in glasses- are they quite different as generally believed in the field, or are they the same as recently suggested by Ingram et al. [18]? The large decoupling index $R_\tau$ observed in glasses generally result because activated volumes for ion-transport are much smaller than for structural relaxation. The elastic interactions that are active in ion-hopping at short and long time scales and that locally deform a network to create a doorway for an ion to hop across, are essentially the same in nature that lead to mechanical relaxation of the network as a whole. In our experiments, the AgI salt provides carriers to the base $AgPO_3$ glass, and it also elastically softens the matrix. As long as the matrix is stiff carriers are localized, however as the matrix softens ion-transport becomes pervasive.

Evidence for channel or filamentary conduction is also found in metal modified silicates. The findings in the silicates have a strong bearing to the behaviour observed in the present electrolyte system. In silicates, it has been found both from Molecular Dynamics simulations [80, 81] and experiments [82, 83] that a new intermediate range order emerges with addition of alkali network modifiers. In silica-rich compositions, local intra-channel hopping prevails leading to high activation barriers and low ionic conductivities. However, with increasing alkali content evidence of clustering is suggested, and at x > 16% these clusters join to create channels in a modified random network. New characteristic metal-metal distances appear as manifested in e.g. pair correlation functions. The composition at which such micro-segregation sets in, turns out to be the elastic phase boundary between stressed-rigid and intermediate phase [84]. In silicates, percolation of channels and onset of conduction are clearly related to the intermediate phase.

The present electrolyte glass system, much like the silver iodomolybdates are examples of strongly decoupled systems, yet one finds that in both these systems conductivity is largely controlled by the elasticity of the network. The case of molten salts such as $2Ca(NO_3)_2.3KNO_3$ or CKN is of interest because these are examples of mildly decoupled electrolytes. In these systems, modelling studies show [85] that frequency dependence of σ can be related to the dynamical properties of melts, again supporting the view that basic interactions that control ion-transport and separately network structure relaxation are essentially the same. The equivalence of the Ngai coupling model [82, 86, 87] with the Funke's concept on mismatch generated relaxation for the accomodation and transport of ions [87] (migration) , reinforces the view that compliance of the disordered matrix plays a central role in determining ion-transport in electrolyte glasses.

## E. Boson and floppy modes in elastically flexible phase

Presence of excess vibrations over Debye-like ones (density of states $g(\omega) \sim \omega^2$) due to a redistribution of the vibrational density of states in glasses over their crystalline counterparts usually gives rise to a feature in the low-frequency range (10-50 $cm^{-1}$) called the boson mode. These excess modes have been observed in Raman scattering [88-90] and in inelastic neutron scattering. The origin of these low-frequency vibrations in a disordered solid has been the subject of ongoing discussions [89, 90]. One view is to regard glasses to be heterogeneous, i.e, to possess a domain structure. Strong intra-domain forces lead atoms to vibrate coherently in a domain. Weak inter-domain forces lead to soft modes associated with transverse acoustic vibrations usually attributed to boson mode [88]. Another source of excess vibrations, particularly in flexible networks, can be floppy modes. These are zero-frequency solutions of the secular determinant and in practice are shifted to finite frequencies due to residual interactions [91]. They have been probed in Lamb-Mossbauer factors [92] and inelastic neutron scattering experiments [93, 94]. For a network of chains, each atom having 2 near neighbors, $\boldsymbol{r}$ = 2, in a 3D network constrained by bond-stretching and bond-bending forces, there is one floppy mode (f = 1) per atom In rigidity theory [95] flexibility of networks is mathematically cast in terms of the count of floppy modes per atom, and when such networks become rigid upon increased cross-linking, the count of floppy modes f($r$) vanishes defining the rigidity percolation threshold when $\boldsymbol{r}$ = 2.40 according to equation (7),

$$f = 6 - \left(\frac{5}{2}\right)\boldsymbol{r} \qquad (7)$$

In such a mean-field description of rigidity, one finds that the slope df/d$r$ = 2.50.

Do floppy modes have a bearing to boson modes in glasses? The observation of the three elastic phases and the availability of rather complete compositional trends in frequency and scattering strength of the boson mode (Figure 8) offer the opportunity to address the basic issue. As mentioned earlier, variations in the boson mode frequency and scattering strength display two thresholds that coincide with the two elastic thresholds, unequivocally showing the elastic behaviour of these glasses must have a direct bearing to boson modes. The linear increase in the bose mode scattering strength $\frac{d\left[\frac{I_{red}(x)}{I_{red}(100)}\right]}{dx} = 1.41$ at x > 37.8% ( (Figure 8) invites the following comment. In 3D networks, rigidity theory has shown [95] that the count of floppy modes per atom (f) decreases linearly with mean coordination number *r*, with a slope $df_{theo}/dr = 2.50$. In the present electrolyte glasses, the connection between glass composition x and its mean coordination number ***r*** is not as obvious as in covalent networks, largely because valence of ions are usually not satisfied locally. We had recognized several years ago [3] that the end member composition, x = 100% or AgI glass is topologically equivalent to a chain of atoms in which each atom has two neighbours, i.e., ***r*** = 2, such as in glassy Se. We now make the simplest assumption [91] that the mean coordination number ***r*** at the onset of rigidity in the present electrolytes (at x = 37.8%) has a value of 2.40. The assumption permits us to directly connect changes in x to those in r. Secondly, we recognize that the observed normalized scattering strength, $\frac{I_{red}(x)}{I_{red}(100)}$ would represent the floppy mode count, since f = 1 at x = 100% when ***r*** = 2. Under these two assumptions, one can show that the observed slope $\frac{d\left[\frac{I_{red}(x)}{I_{red}(100)}\right]}{dx} = 1.41$, translates to a $\frac{df_{obs}(x)}{d\boldsymbol{r}}$ of 2.30(3), which may be compared to the rigidity theory predicted slope $\frac{df_{theo}(x)}{d\boldsymbol{r}}$ of 2.50. These data strongly suggest that the observed boson mode in the flexible phase (at x > 37.8%) of the present glasses represents floppy mode excitations. The most natural interpretation of the present data can be schematically illustrated by the plot shown in Figure 8b. In the flexible glasses, floppy modes largely contribute to the boson mode as shown by the broken line with a slope $\frac{df_{theo}(x)}{dr}$ = 2.50. And as glasses enter the intermediate phase at x < 37.8%, soft modes contribute contribute to boson mode. This particular feature of a growth in the soft mode scattering in the intermediate phase is not well understood at present, but we note that earlier results on $Ge_xSe_{1-x}$ glasses are consistent with that finding [44.]

The identification of floppy modes contributing to low-frequency modes in Raman scattering of the present electrolyte glass system is profound for another reason. To date floppy modes have been identified in $Se_n$-chain bearing covalent networks as arising from the lack of dihedral angle restoring forces [96]. In the present electrolyte glasses, there are no 2-fold coordinated chains, but the system configurationally evolves in an identical fashion. For a AgI glass, there is one floppy mode per atom that contributes to its flexibility. The microscopic origin of this mode is at present not entirely obvious.

In summary, the present findings on low-frequency vibrations observed in Raman scattering on a prototypical solid electrolyte glass show that the so called boson mode have contributions from both soft modes and floppy modes, with the former contributing in the mildly rigid elastic phase and the latter in the flexible elastic phase. Ideas on elastic behaviour of glasses common place in covalent solids, apparently also extend to ionic solids including fast-ion conducting systems as well.

## V. CONCLUSIONS

In the present work we have demonstrated the existence of three distinct regimes of elastic behaviour in dry $(AgI)_x(AgPO_3)_{1-x}$ glasses, 0 < x < 9.5% stressed-rigid, 9.5 < x <

37.8 % Intermediate, x > 37.8 % flexible. Raman optical elasticity power-laws, trends in the nature of the glass transition endotherm corroborate the three elastic phase assignments. Ionic conductivity measurements reveal a step-like increase when glasses become stress-free at $x > x_c(1) = 9.5\%$, and a logarithmic increase in conductivity once glasses become flexible at x > xc(2) = 37.8 % in the dry samples with a power-law t = 1.75. The power-law is suggestive that the logarithmic increase results due to percolation of 3D filamentary conduction pathways. These data represent the intrinsic behaviour of dry glasses, which are in contrast to earlier reports on wet glasses synthesized by handling precursors at laboratory ambient environment, which reveal a monotonic increase in conductivity with the solid electrolyte fraction but no steps. These data demonstrate rather directly the central role of network flexibility in controlling ion-transport in a prototypical decoupled glass.

The widely different relaxation times for ion-transport and structure relaxation in glasses, derives from the underlying size of volumes impacted by these processes, which is minuscule (less than $1A^3$) in the former and global (~ $cm^3$) in the latter. However, in both these cases process compliance or elasticity has the same magnitude as illustrated by the unified approach of Ingram et al.[18]. These ideas bring glasses and polymer electrolytes on the same platform, and are in harmony with the present finding that elastic flexibility of solid electrolyte backbones promotes ion-transport.

Variations of boson mode frequency and scattering strength display two thresholds that coincide with the two elastic phase boundaries. In the flexible phase, scattering strength of the boson mode increases almost linearly with glass composition x, with a slope that tracks the *floppy mode fraction* as a function of mean coordination number ***r*** predicted by mean-field rigidity theory**.** These data unequivocally show that the excess low frequency vibrations contributing to boson mode in flexible glasses must come in large part from *floppy modes*.

**Acknowledgements**

It is a pleasure to acknowledge discussions with Professor Malcolm Ingram, Professor Bernard Goodman and Professor Eugene Duval. This work is supported by NSF grant DMR 04-56472.

**Table 1.** Vibrational mode assignments of the current work along with suitable references to previous work in the field.

**Figure 1.** (Color online) Variations in molar volume $V_m(x)$ in set A,(dry:▼), set B (wet :▲)$(AgI)_x(AgPO_3)_{1-x}$ glasses synthesized in present work, and those reported by Sidebottom (■) in ref. 26.

**Figure 2.** X-ray powder diffraction pattern of $AgPO_3$ glass sample in set A. Note that the most intense Bragg peak in XRD results of our sample occurs at 2Θ = 29.30 deg.. It is shifted to a lower value than the published result (2Θ = 31.67deg from the Joint Committee on Powder Diffraction Standards (JCPDS) [39] data file # 11-0640.

**Figure 3.** (Color online) MDSC scans of present $(AgI)_x(AgPO_3)_{1-x}$ glasses (set A) at (a) x = 5% (Stressed-Rigid), (b) x = 17 % (Intermediate) and (c) x = 40% (Flexible). Note that non-reversing enthalpy at x = 5% sample displays a broad peak (FWHM, W = 35(2) °C) that is asymmetric (high-T tail). At x = 17% the width W=25(2) °C, and the peak becomes symmetric. At x = 50% the width W decreases to 15(2) °C and the peak remains symmetric. These trends in the width W and shape of the non-reversing heat flow observed in the present glasses are quite similar to those seen earlier in covalent glasses.

**Figure 4.** (Color online) Variations in (a) $T_g(x)$, (b) non-reversing enthalpy $\Delta H_{nr}(x)$ (c) the variation in heat capacity change at $T_g$, $\Delta C_p(x)$, in set A dry (▼), set B wet (▲) $(AgI)_x(AgPO_3)_{1-x}$ glass samples synthesized in present work, and those reported by Mangion-Johari (●) ref. 24, Sidebottom (■) ref. 26 and Hallbrucker-Johari (●) ref. 22. Note that presence of water in glass samples reduces $T_g$ of the base material ($AgPO_3$) and also width of the reversibility window ( compare set A with set B).

**Figure 5.** (Color online) Raman scattering on $(AgI)_x(AgPO_3)_{1-x}$ glass samples A in the low (0-400 $cm^{-1}$) frequency range. The observed lineshapes are analyzed in terms of three modes; a mode centered near 40 $cm^{-1}$ ( boson mode), a $Ag^+$ restrahlen mode near 85 $cm^{-1}$, and a mode of AgI near 130 $cm^{-1}$. Scattering strength of these modes increases with AgI content (figure 6).

**Figure 6**. (Color online) Raman lineshapes of present $(AgPO_3)_{1-X}(AgI)_X$ glass samples A at several different compositions. Broadly, these results suggest glass structure to transform from chain-like to ring-like as the AgI content increases.

**Figure 7.** (Color online) Reduced Raman scattering, $I_{red} = \dfrac{I_{\exp}}{n_B + 1} \propto \dfrac{C(\omega)g(\omega)}{\omega}$, of $(AgI)_x(AgPO_3)_{1-x}$ glass samples A at different glass compositions in the low (< 200 $cm^{-1}$) frequency range. $I_{exp}$ represent the experimental Raman scattering intensity, C(ω) the photon-vibration coupling constant [46], g(ω) the vibrational density of states (VDOS), and $n_B$ the Bose occupation number. The inset figure shows the deconvolution of reduced Raman spectra of $(AgI)_x(AgPO_3)_{1-x}$ at x = 40% in terms of three modes as mentioned earlier in Figure 5.

**Figure 8.** (Color online) Variations in (a) the frequency and (b) scattering strength of the boson mode of $(AgI)_x(AgPO_3)_{1-x}$ glass samples A as a function of AgI content. Note that the results reveal thresholds near the stress ($x_c(1)$ = 9.5%) and rigidity ($x = x_c(2)$ = 37.8%).

**Figure 9.** (Color online) Infrared absorption of $(AgPO_3)_{1-X}(AgI)_X$ glass samples A at several different compositions.

**Figure 10.** Variations in (a) the P-$O_t$ symmetric mode frequency in Raman scattering, and (b) the P-$O_t$ asymmetric mode frequency in infrared absorption experiments as a function of AgI. Note that the P-$O_t$ symmetric and asymmetric modes systematically red-shift with increasing x to display two thresholds, one near $x_c(1)$ = 9.5% and a second near $x_c(2)$ = 37.8%. .

**Figure 11.** (a) shows a plot of $\log_{10}(\nu^2 - \nu_c(1)^2)$ against $\log_{10}(x_c(1) - x)$ in the $0 < x <$ 9.5% range, and gives the optical elastic power law in the stressed rigid glass samples A, $p_1$ = 1.25(2). (c) Shows a plot of $\log_{10}(\nu^2 - \nu_c(2)^2)$ against $\log_{10}(x_c(2) - x)$ in the 9.5% $< x <$ 37.8% range and gives the optical elastic power law in the intermediate phase of glass samples A, $p_2$ = 0.98(3).

**Figure 12.** (Color online) Variations in electrical conductivity σ(x) (▼), and activation energy for conductivity $E_A(x)$ (●) of present $(AgPO_3)_{1-X}(AgI)_X$ glass samples A with glass composition. The frequency dependence of conductivity permits fixing the high

frequency permittivity and the Coulomb Energy $E_c$(♦). The resulting variations in the strain energy (▲)$E_s = E_A - E_c$ , shows the term to remain high in the stressed-rigid but to decrease in the flexible phase.

**Figure 13.** (Color online) Variations in room temperature conductivities, σ(x) in (a) present glass samples A (▼), and (b) those reported by Mangion-Johari [23, 24] (● and ★), Sidebottom [26] (■), Bhattacharya et al.[54] (♦ ), and Malugani et al.[61](✥)

**Figure 14.** Shows a plot of $\log_{10} \sigma$ against $\log_{10} (x - x_c(2))$ and yields a conductivity power-law, $\mu = 1.78(10)$ in the flexible phase, with $x_c(2) = 37.8\%$ in present glass samples A.

**TABLE I**: $AgPO_3$ Mode assignments

| Mode Assignment | Present Work Raman ($cm^{-1}$) | Present Work Infrared ($cm^{-1}$) | References |
|---|---|---|---|
| **Boson** | 28.8 | - | Fontana et al. [49, 67] |
| **$Ag^+$ ions oscillations inside their oxygen cage** | 85 | 51.1<br>139.7 | Mercier et al. [38]<br>Kamitsos et al. [48] |
| **$\nu_s$ (Ag-I)** | 130 - 120 | 130 - 120 | Mercier et al. [38]<br>Kamitsos et al. [48]<br>Shastry & Rao [51] |
| **Bond bending modes of different structural units** | 650-300 | 650-300 | • Kamitsos et al. [48] & Velli et al. [49] Suggested that mode at 650 - 400 $cm^{-1}$ is composed of bending modes of different structural units.<br>• Nelson & Exarhos [50] suggested that mode at 360 - 300 $cm^{-1}$ is composed of pendent and in-chain P-Ot bending modes. |
| **$\nu_s$ ($P\text{-}O_{br}$) long chain** | 686.6 | 685.9 | Shastry & Rao [51]<br>Rulmont et al. [59]<br>Kamitsos et al. [48] |
| **$\nu_s$ ($P\text{-}O_{br}$) large ring** | 766.3 & 733.7 | 773 & 731 | Rulmont et al. [59] |
| **$\nu_{as}$ ($P\text{-}O_{br}$) long chain** | 900 | 894 | Rulmont et al. [59]<br>Kamitsos et al. [48] |
| **$\nu_{as}$ ($P\text{-}O_{br}$) large ring** | 920.3 | 917.3 | Rulmont et al. [59]<br>Kamitsos et al. [48] |
| **$\nu_{as}$ ($P\text{-}O_{br}$) small ring** | 971 | 969.4 | Rulmont et al. [59]<br>Kamitsos et al. [48] |
| **$\nu_s$ ($P\text{-}O_{br}$) small ring** | 1008 | 1010.5 | Rulmont et al. [59]<br>Kamitsos et al. [48] |
| **$\nu_{as}$ ($PO_3^{2-}$)** | 1087 | 1103.8 | Kamitsos et al. [48]<br>Velli et al. [49] |
| **$\nu_s$ ($PO_3^{2-}$)** | 1029 | - | Rulmont [59] |
| **$\nu_s$ ($P\text{-}O_t$) long chain** | 1140 | 1162.5 | Shastry & Rao [51]<br>Rulmont et al. [59]<br>Kamitsos et al. [48] |
| **$\nu_{as}$ ($P\text{-}O_t$) long chain** | 1235.5 | 1256.2 | Rulmont et al. [59]<br>Kamitsos et al. [48] |

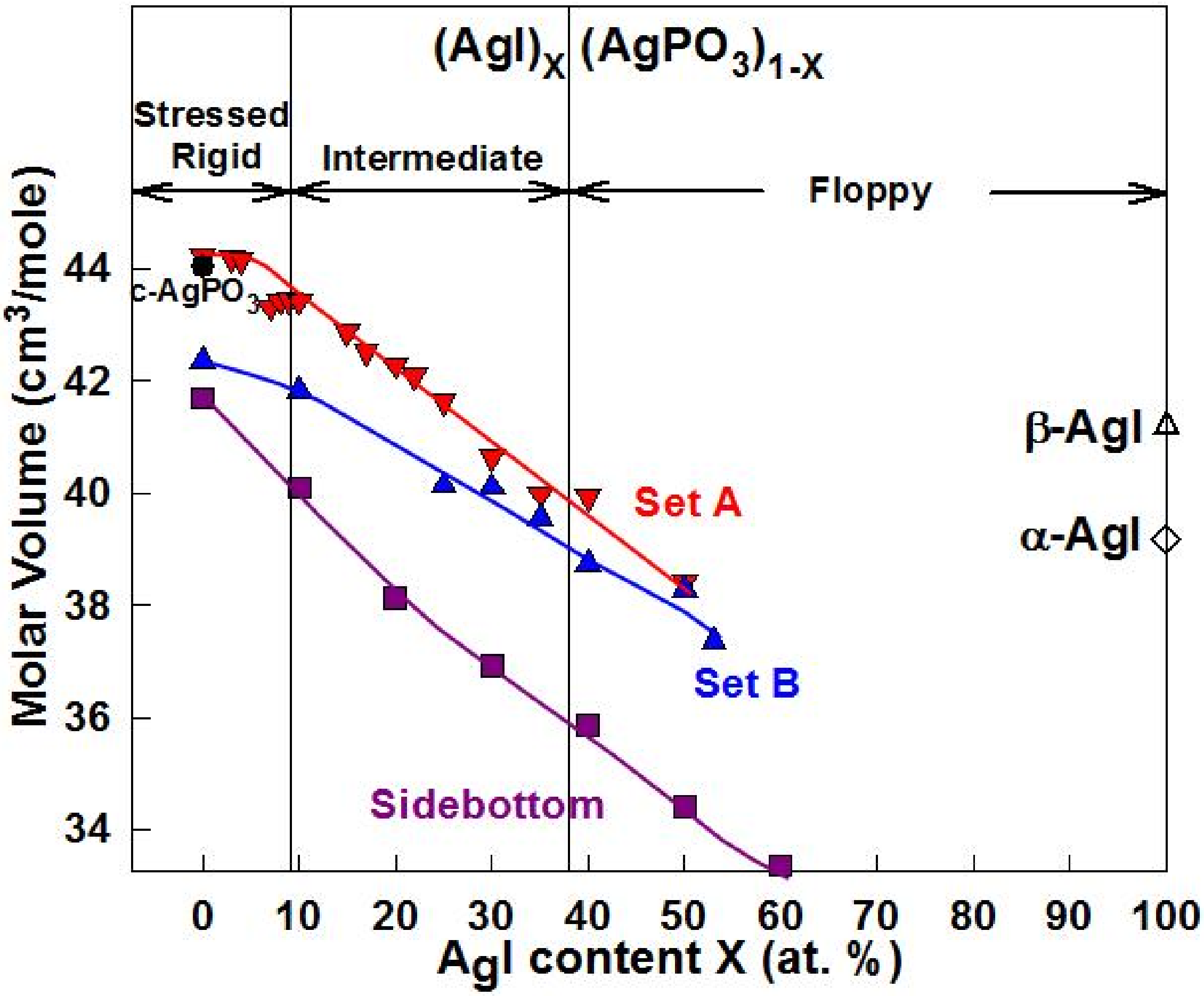

**Figure 1**

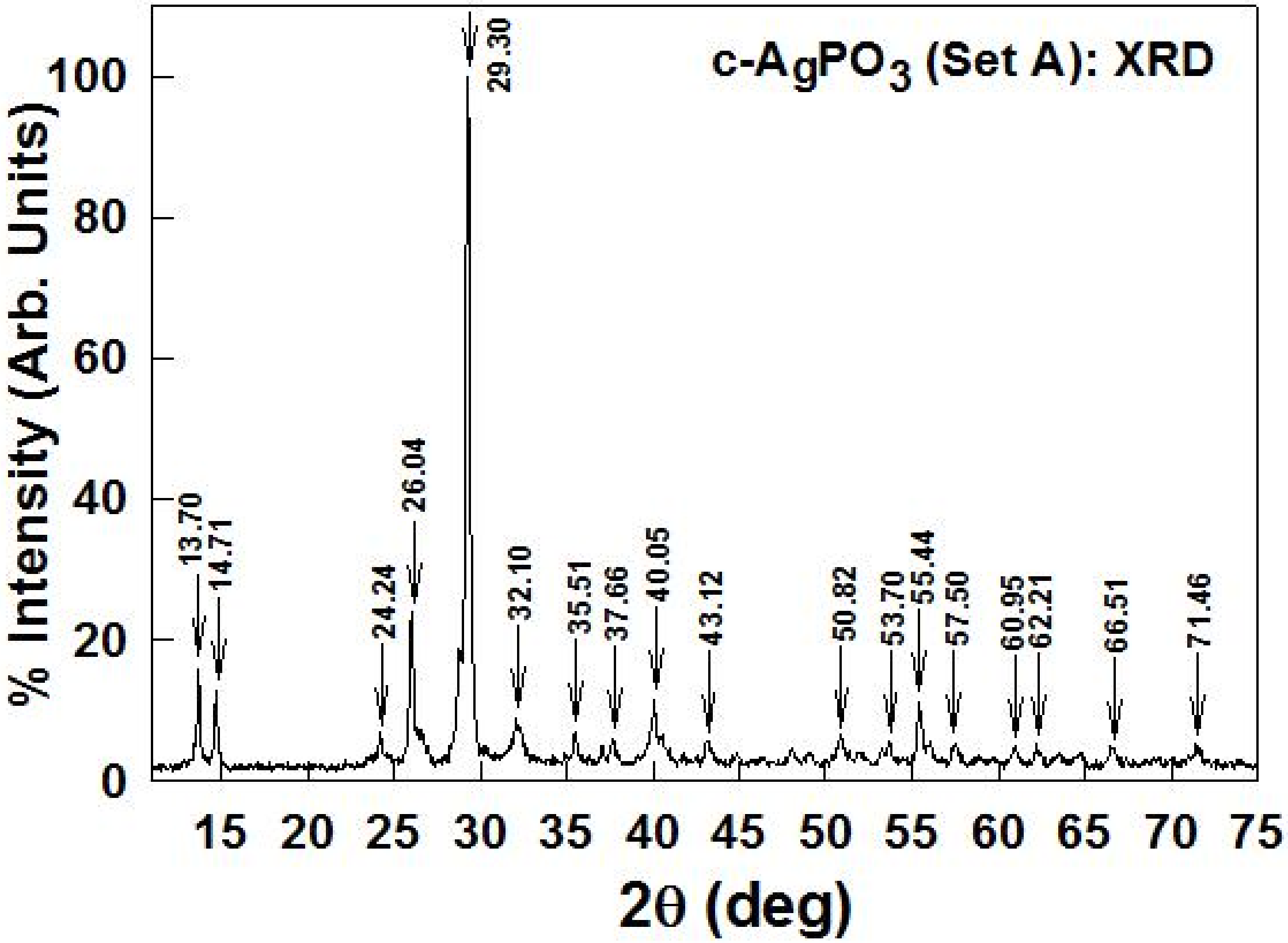

**Figure 2**

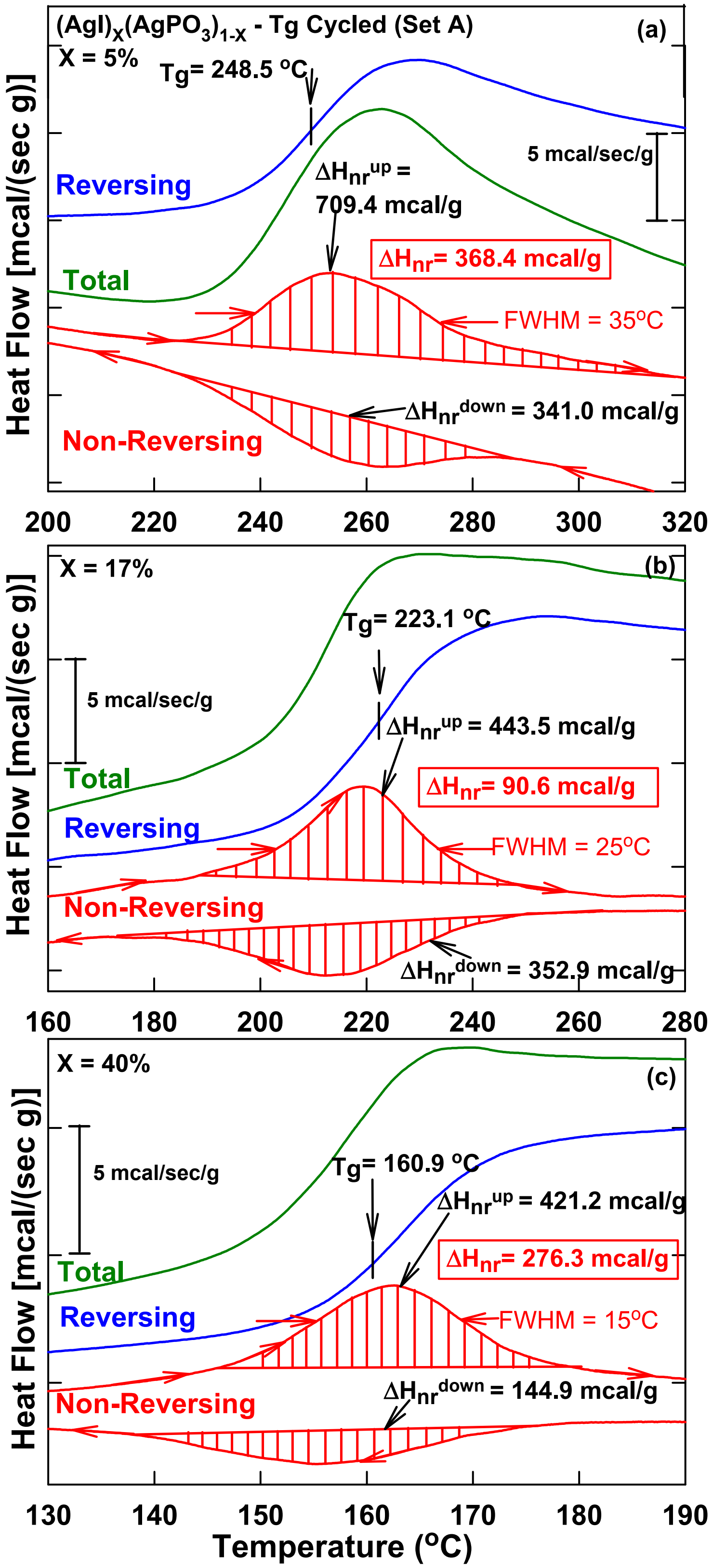

(AgI)X(AgPO3)1-X - Tg Cycled (Set A)
(a)
X = 5%
Tg= 248.5 °C
5 mcal/sec/g
Reversing
ΔHnrup = 709.4 mcal/g
ΔHnr= 368.4 mcal/g
Total
FWHM = 35°C
ΔHnrdown = 341.0 mcal/g
Non-Reversing
200 220 240 260 280 300 320
X = 17%
(b)
Tg= 223.1 °C
5 mcal/sec/g
ΔHnrup = 443.5 mcal/g
Total
ΔHnr= 90.6 mcal/g
Reversing
FWHM = 25°C
Non-Reversing
ΔHnrdown = 352.9 mcal/g
160 180 200 220 240 260 280
X = 40%
(c)
5 mcal/sec/g
Tg= 160.9 °C
ΔHnrup = 421.2 mcal/g
ΔHnr= 276.3 mcal/g
Total
Reversing
FWHM = 15°C
ΔHnrdown = 144.9 mcal/g
Non-Reversing
130 140 150 160 170 180 190
Heat Flow [mcal/(sec g)]
Temperature (°C)

**Figure 3**

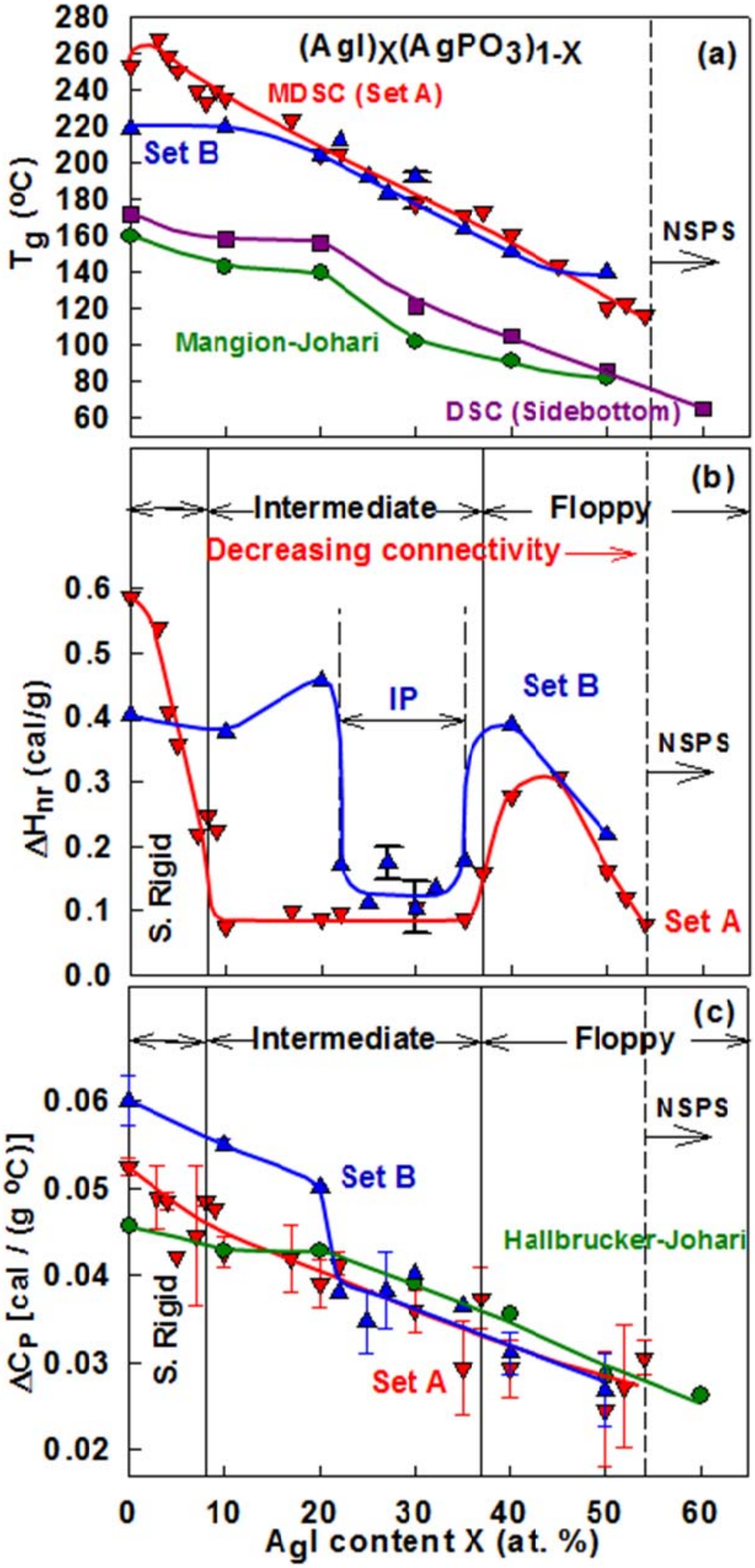

**Figure 4**

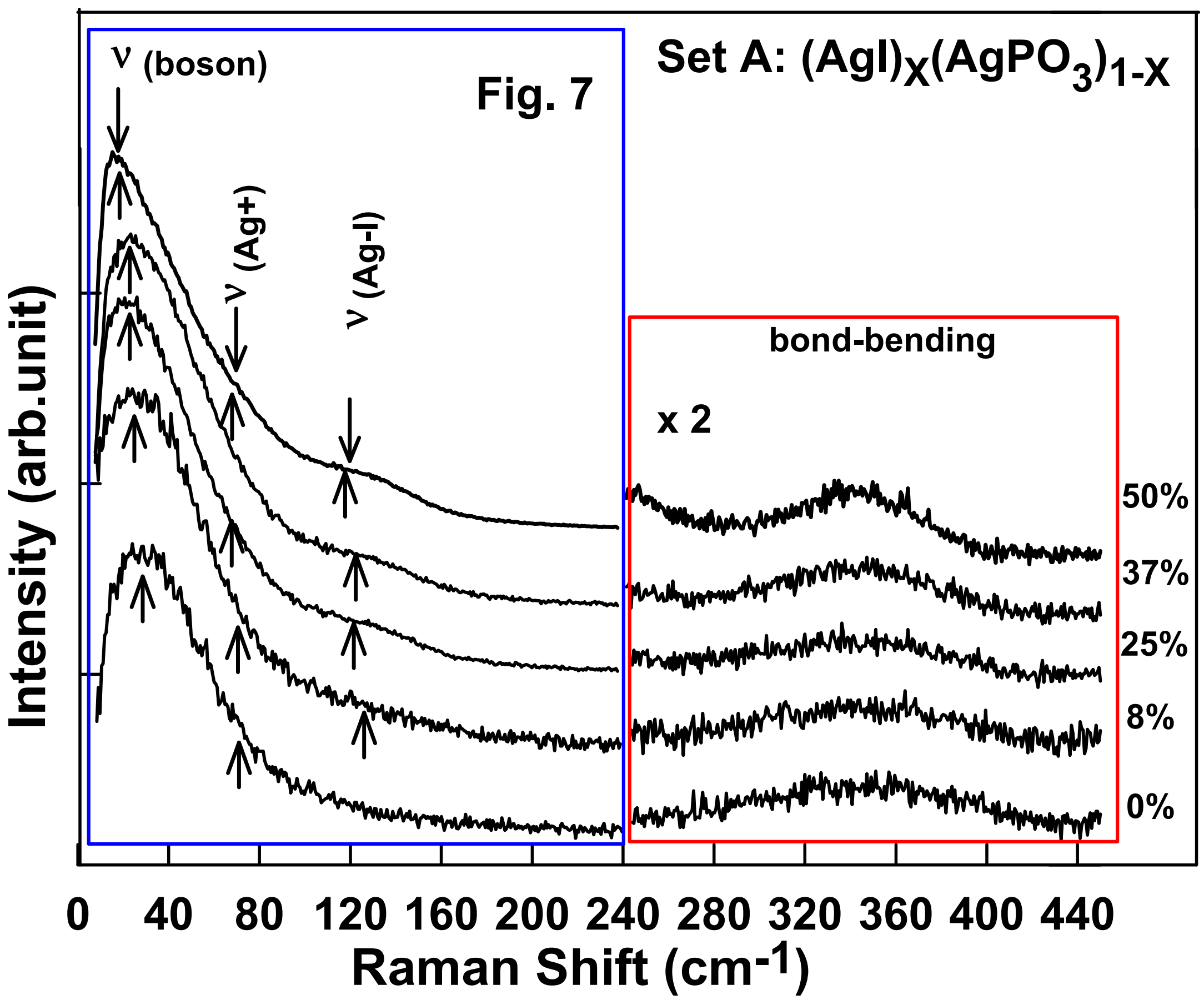

**Figure 5**

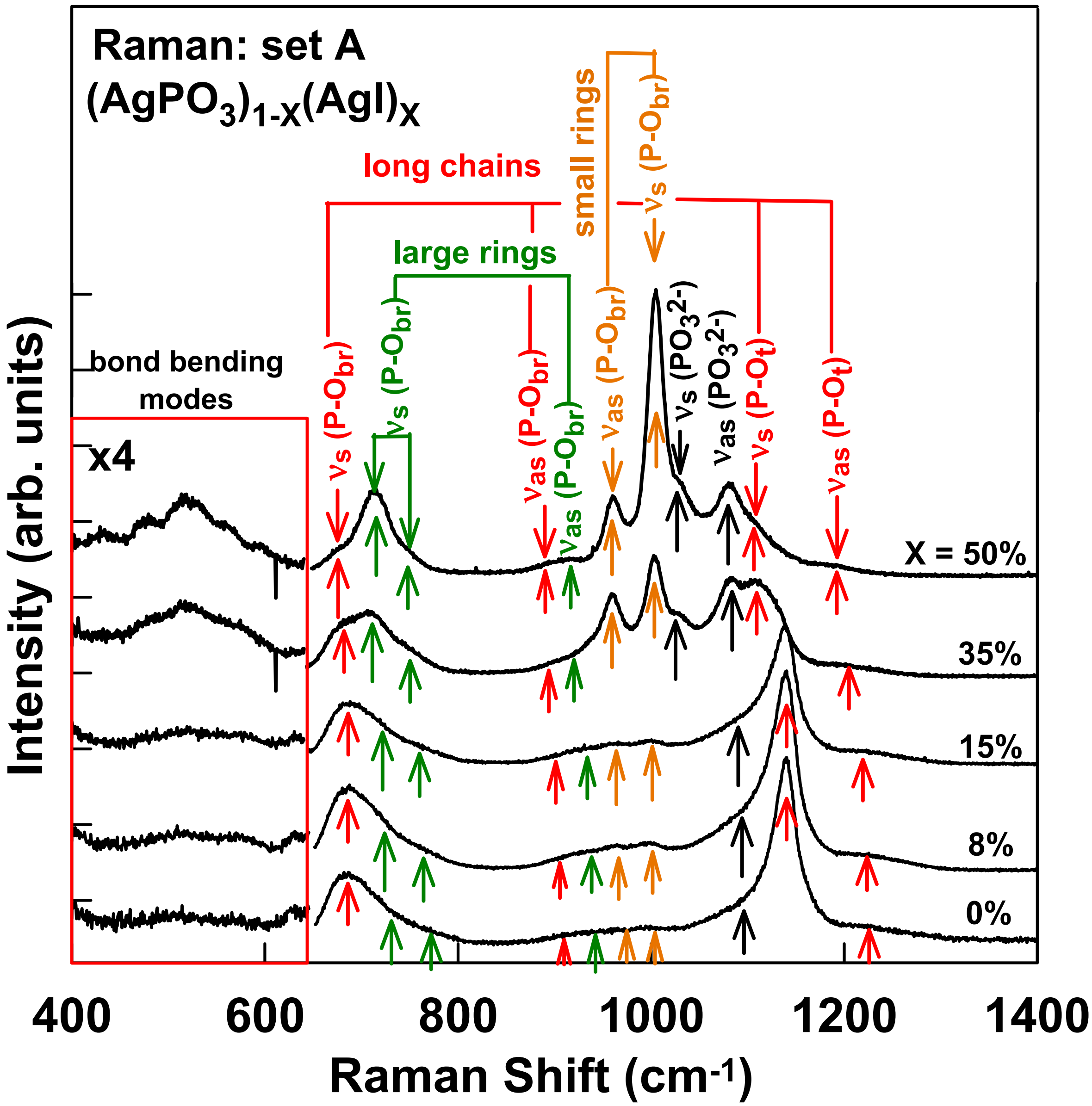

**Figure 6**

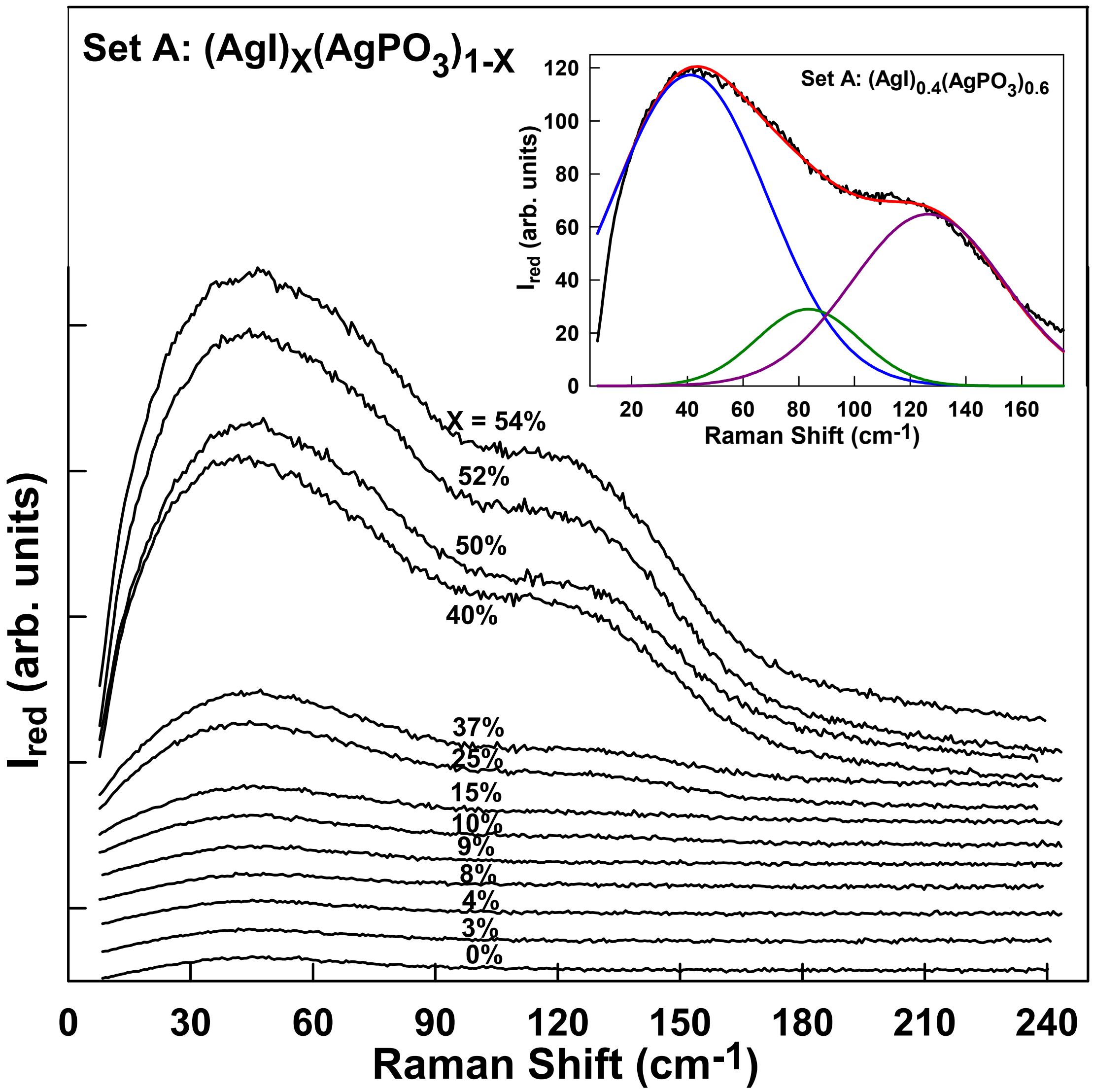

**Figure 7**

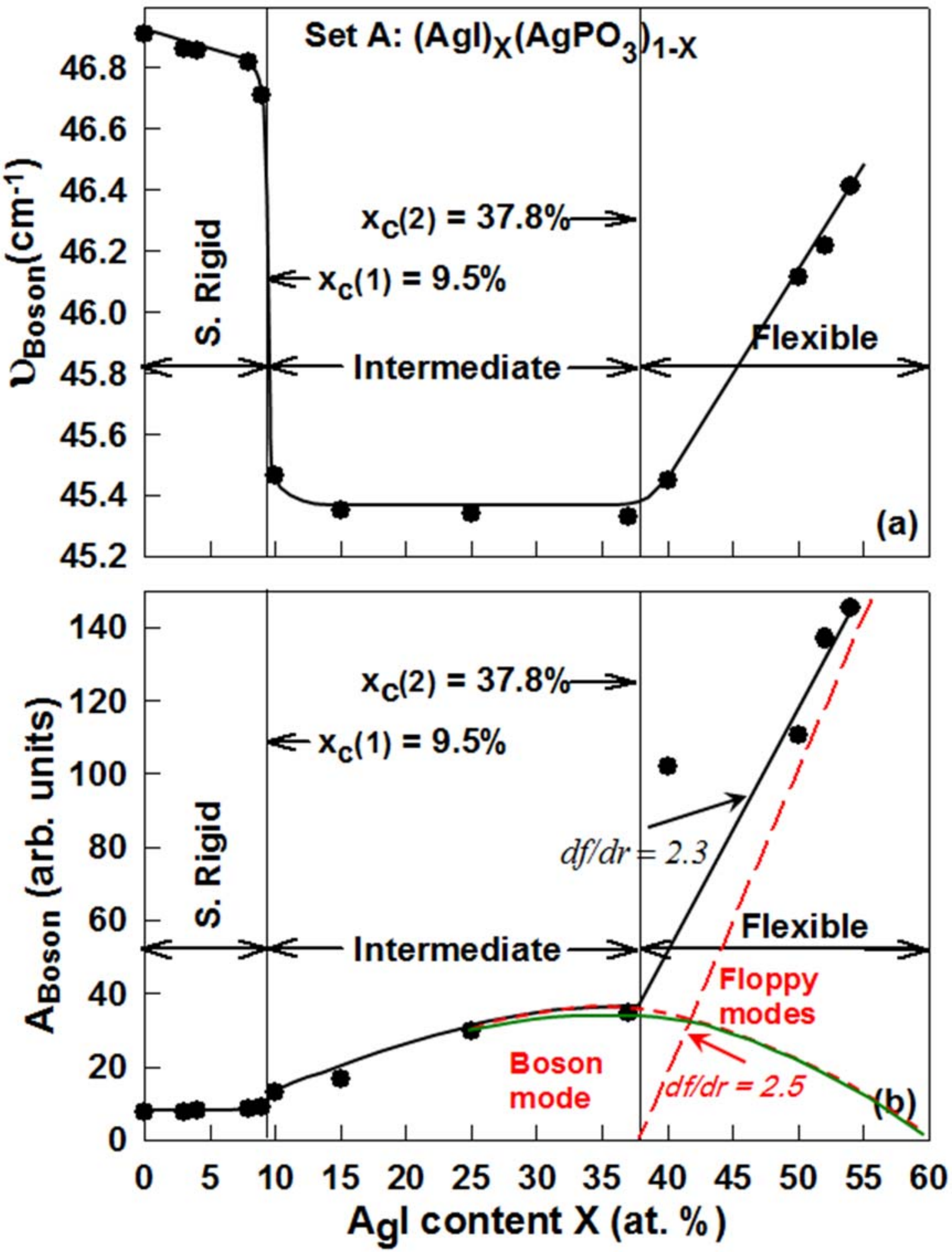

**Figure 8**

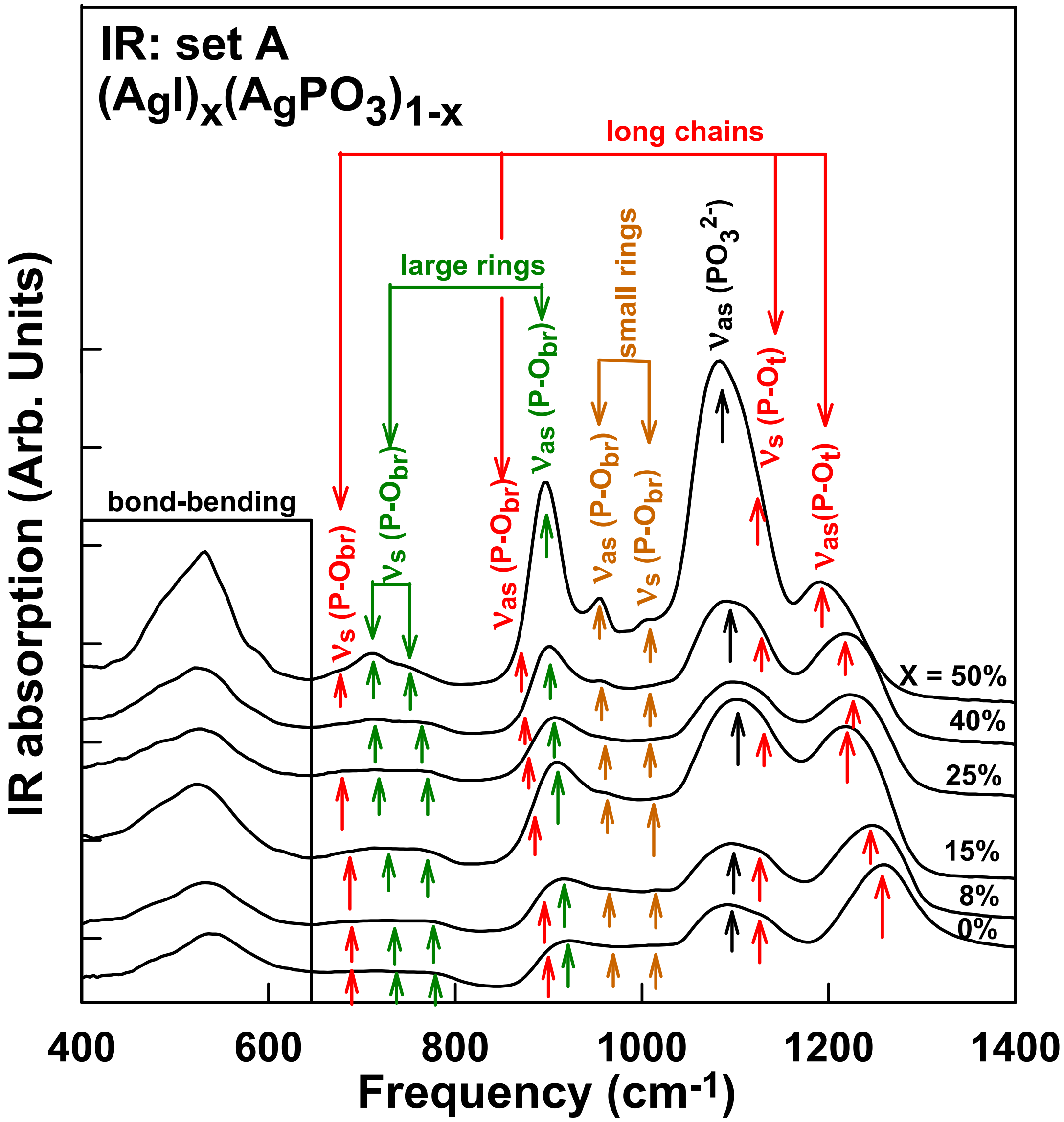

**Figure 9**

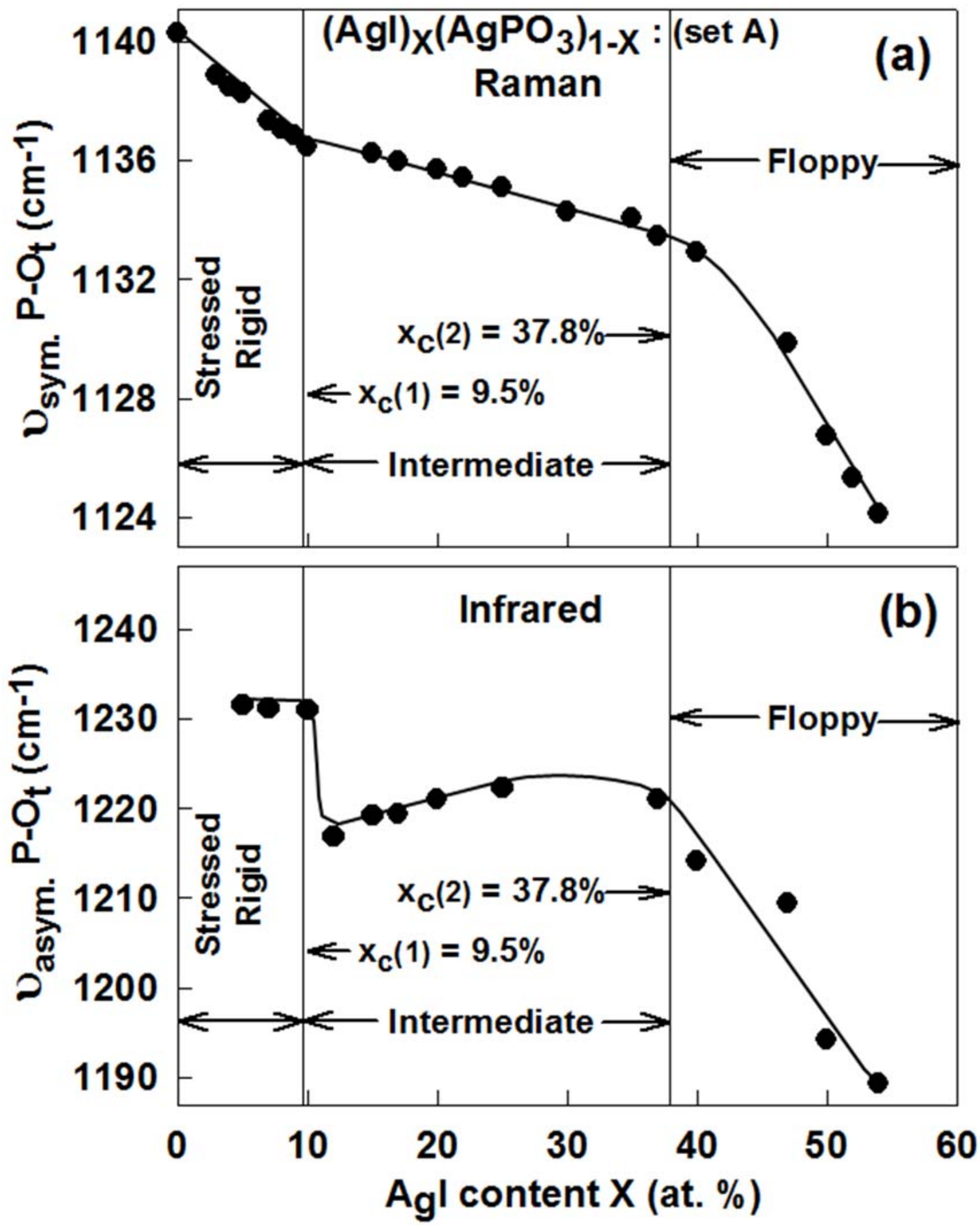

**Figure 10**

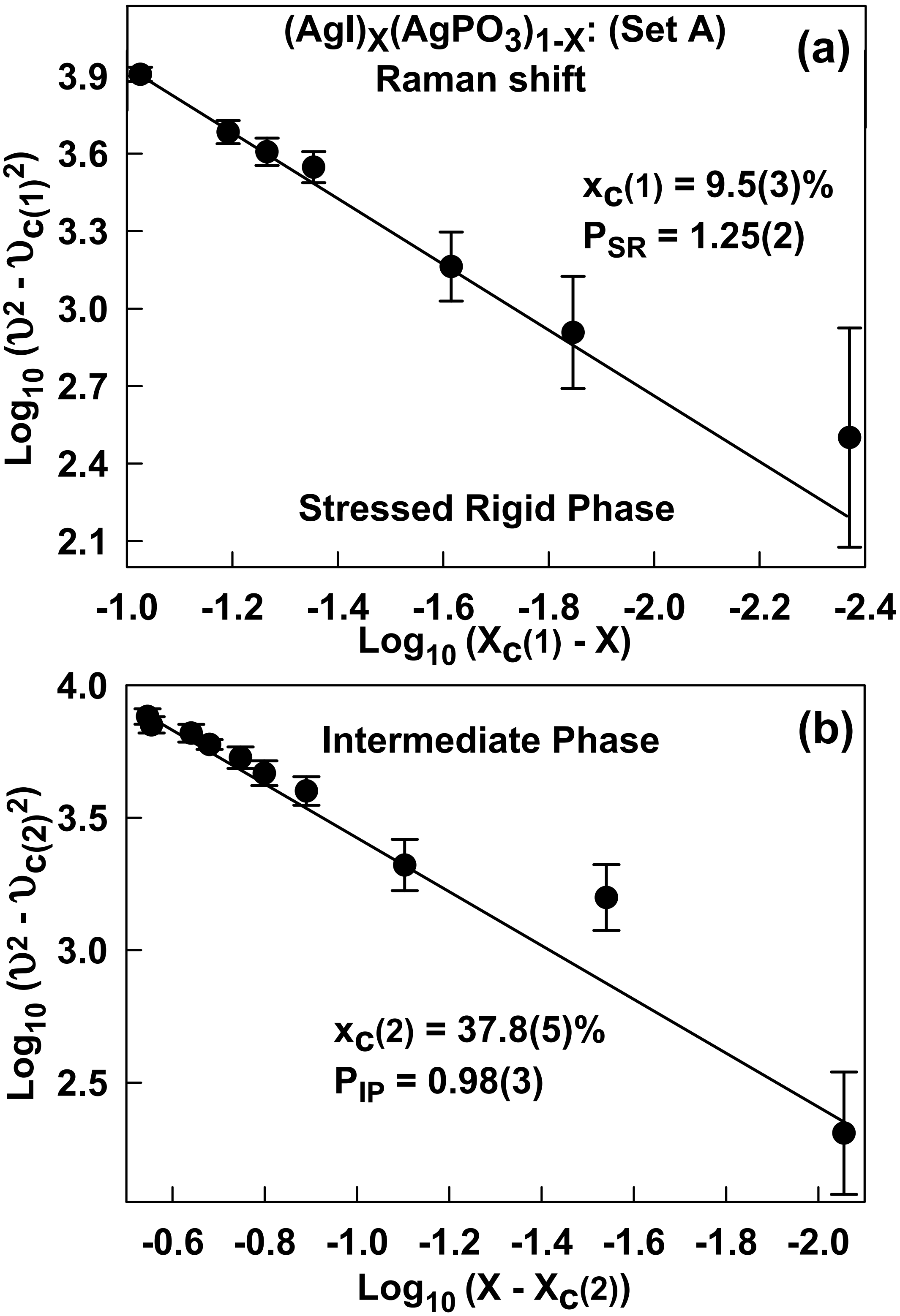

Figure 11

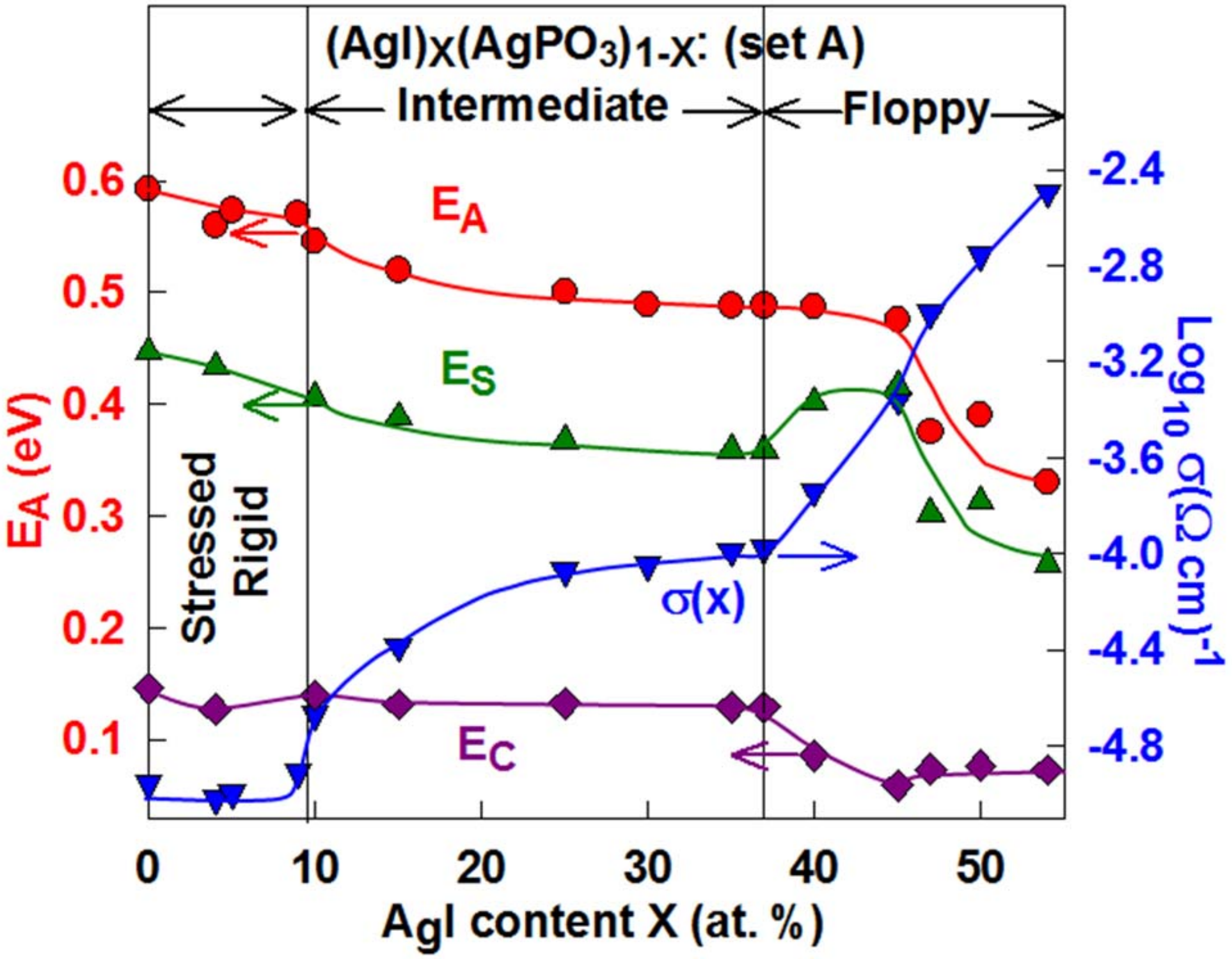

**Figure 12**

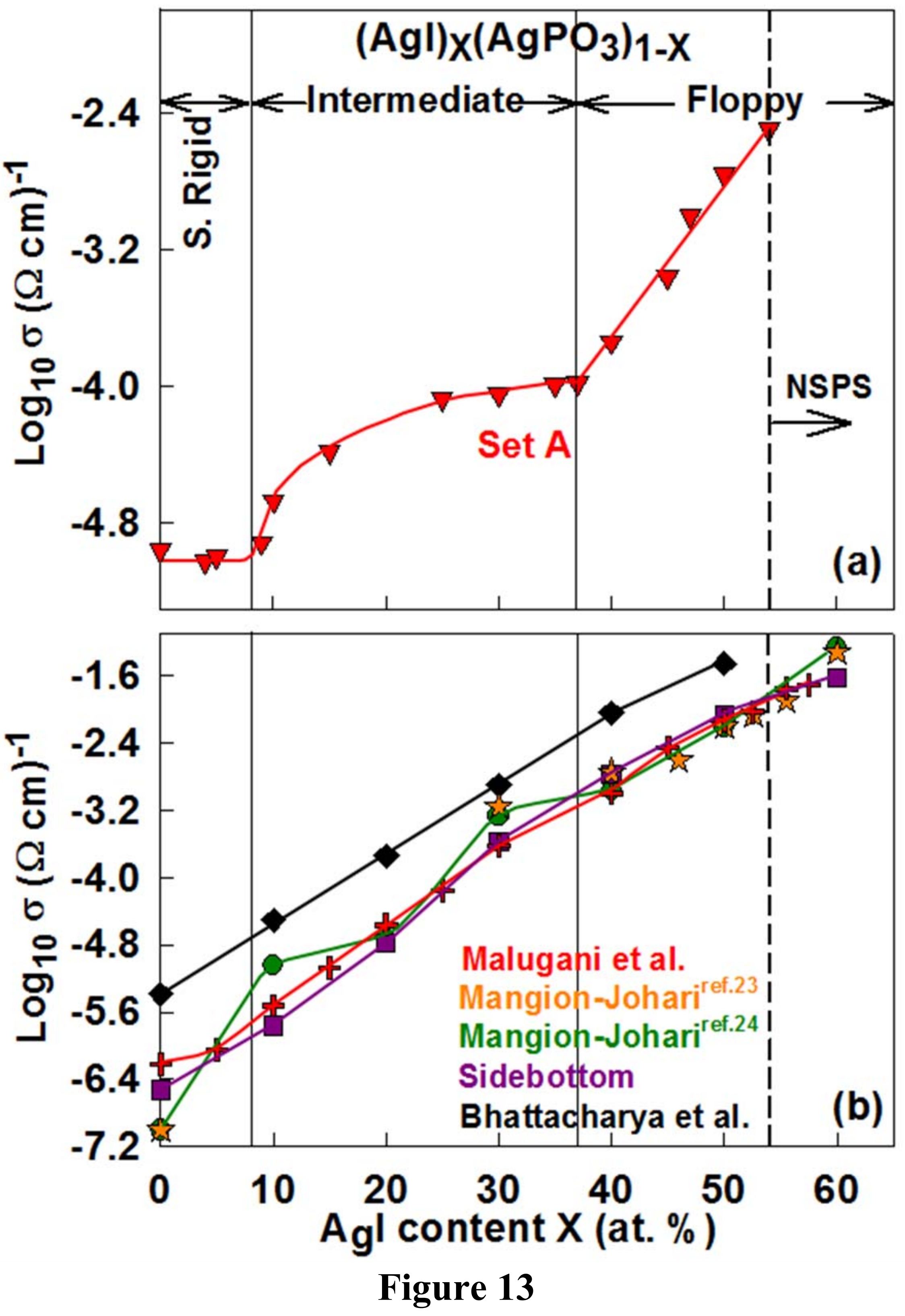

**Figure 13**

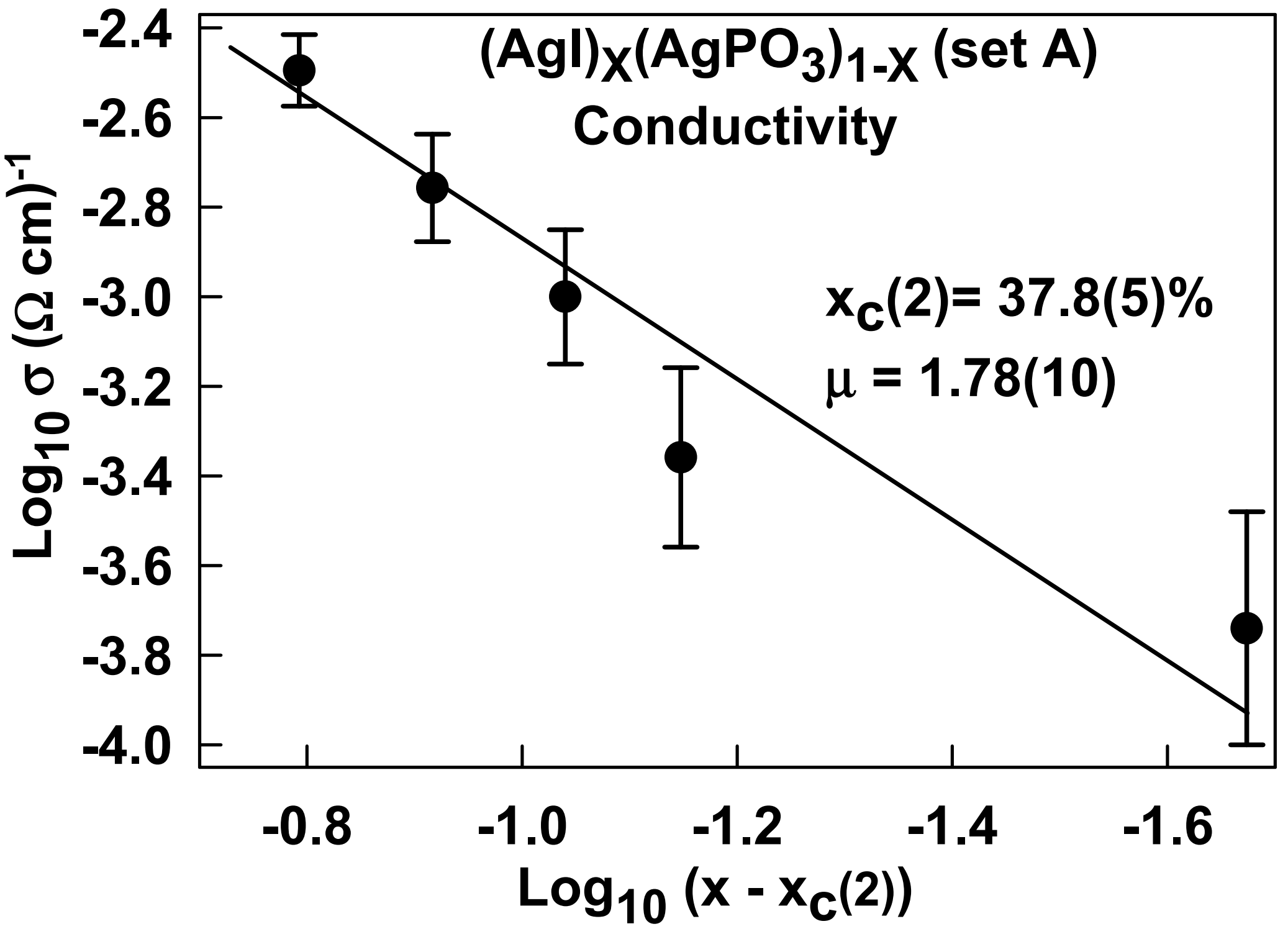

**Figure 14**